\documentclass[preprint,showpacs,aps,draft]{revtex4}

\usepackage{graphicx}
\usepackage{dcolumn}
\usepackage{bm}

\input epsf
\newcounter{subequation}[equation]
\makeatletter
\fussy
\def \with respect to { with respect to }

\def\l {\lambda } 
 
\def \t {\theta }

\def\a {\alpha }
\def\dh {\partial }
\def \d {\delta }
\def \D  {\Delta }

\def \g {\gamma }

\def \b {\beta }

\def \e {\epsilon }

\def \ud { {1 \over 2} }

\def \cala { {\cal A } }

\def \calr { {\cal R } }
\def \cals { {\cal S } }

\def \cals { {\cal S } }
\def\vb#1{\vbox to #1 pt{}}
\def \tchi  {{\tilde \chi }}
\def \Eslash {E \kern-.5em\slash}
\def \pslash {p \kern-.5em\slash}
\def \kslash {k \kern-.5em\slash}
\def \Dslash {D \kern-.5em\slash}
\def \hslash {h \kern-.5em\slash}
\def \dslash {\partial \kern-.5em\slash}
\def \vslash {v \kern-.5em\slash}

\makeatletter

\def\thesubequation{\theequation\@alph\c@subequation}
\def\@subeqnnum{{\rm (\thesubequation)}}
\def\slabel#1{\@bsphack\if@filesw {\let\thepage\relax
   \xdef\@gtempa{\write\@auxout{\string
      \newlabel{#1}{{\thesubequation}{\thepage}}}}}\@gtempa
   \if@nobreak \ifvmode\nobreak\fi\fi\fi\@esphack}
\def\subeqnarray{\stepcounter{equation}
\let\@currentlabel=\theequation\global\c@subequation\@ne
\global\@eqnswtrue
\global\@eqcnt\z@\tabskip\@centering\let\\=\@subeqncr
$$\halign to \displaywidth\bgroup\@eqnsel\hskip\@centering
  $\displaystyle\tabskip\z@{##}$&\global\@eqcnt\@ne
  \hskip 2\arraycolsep \hfil${##}$\hfil
  &\global\@eqcnt\tw@ \hskip 2\arraycolsep
  $\displaystyle\tabskip\z@{##}$\hfil
   \tabskip\@centering&\llap{##}\tabskip\z@\cr}
\def\endsubeqnarray{\@@subeqncr\egroup
                     $$\global\@ignoretrue}
\def\@subeqncr{{\ifnum0=`}\fi\@ifstar{\global\@eqpen\@M
    \@ysubeqncr}{\global\@eqpen\interdisplaylinepenalty \@ysubeqncr}}
\def\@ysubeqncr{\@ifnextchar [{\@xsubeqncr}{\@xsubeqncr[\z@]}}
\def\@xsubeqncr[#1]{\ifnum0=`{\fi}\@@subeqncr
   \noalign{\penalty\@eqpen\vskip\jot\vskip #1\relax}}
\def\@@subeqncr{\let\@tempa\relax
    \ifcase\@eqcnt \def\@tempa{& & &}\or \def\@tempa{& &}
      \else \def\@tempa{&}\fi
     \@tempa \if@eqnsw\@subeqnnum\refstepcounter{subequation}\fi
     \global\@eqnswtrue\global\@eqcnt\z@\cr}
\let\@ssubeqncr=\@subeqncr
\@namedef{subeqnarray*}{\def\@subeqncr{\nonumber\@ssubeqncr}\subeqnarray}
\@namedef{endsubeqnarray*}{\global\advance\c@equation\m@ne%
                           \nonumber\endsubeqnarray}

\@addtoreset{equation}{section}
\renewcommand{\theequation}{\thesection.\arabic{equation}}
\newcommand{\be}{\begin{equation}} 
\newcommand{\ee}{\end{equation}} 
\newcommand{\ba}{\begin{array}}
\newcommand{\ea}{\end{array}}
\newcommand{\bea}{\begin{eqnarray}} 
\newcommand{\eea}{\end{eqnarray}} 
\newcommand{\bsea}{\begin{subeqnarray}} 
\newcommand{\esea}{\end{subeqnarray}}

\begin{document}
\begin{flushright}
Saclay Preprint  SPhT T06/002 \\
hep-ph/yymmxxx
\end{flushright}
\title{ Nonminimal supersymmetric standard model with lepton number 
violation} \thanks {\it Supported by the Laboratoire de la Direction
des Sciences de la Mati\`ere du Commissariat \`a l'Energie Atomique }
\author{ M. Chemtob} \email{ chemtob@spht.saclay.cea.fr} \affiliation{
Service de Physique Th\'eorique, CEA-Saclay F-91191 Gif-sur-Yvette
Cedex France }
\author{ P.N. Pandita} \email{ ppandita@nehu.ac.in } \affiliation{
Service de Physique Th\'eorique, CEA-Saclay F-91191 Gif-sur-Yvette
Cedex France and\\
Department of Physics, North Eastern Hill University, Shillong 793
022, India \footnote{Permanent address} }
                                                                                
\received{\today}



\begin{abstract} \vspace*{10pt}
We carry out a detailed analysis of the nonminimal supersymmetric
standard model with lepton number violation. The model contains a
unique trilinear lepton number violating term in the superpotential
which can give rise to neutrino masses at the tree level. We search
for the gauged discrete symmetries realized by cyclic groups $Z_N$
which preserve the structure of the associated trilinear
superpotential of this model, and which satisfy the constraints of the
anomaly cancellation. The implications of this trilinear lepton number
violating term in the superpotential and the associated soft
supersymmetry breaking term on the phenomenology of the light neutrino
masses and mixing is studied in detail. We evaluate the tree and loop
level contributions to the neutrino mass matrix in this model.  We
search for possible suppression mechanism which could explain large
hierarchies and maximal mixing angles.

\end{abstract}

\maketitle
                                                                                

\section{INTRODUCTION}

Supersymmetry~\cite{wess} is at present the only known framework in
which the Higgs sector of the Standard Model~(SM), so crucial for its
internal consistency, is natural. A much favored implementation of the
idea of supersymmetry at low energies is the minimal supersymmetric
standard model~(MSSM), which is obtained by doubling the number of
states of SM, and introducing a second Higgs doublet (with opposite
hypercharge to the Standard Model Higgs doublet) to generate masses
for all the SM fermions and to cancel the triangle gauge anomalies.
However, the MSSM suffers from the so-called $\mu$ problem associated
with the bilinear term connecting the two Higgs doublet superfields
$H_u$ and $H_d$ in the superpotential.  An elegant solution to this
problem is to postulate the existence of a chiral electroweak gauge
singlet superfield $S$, and couple it to the two Higgs doublet
superfields $H_u$ and $H_d$ via a dimensionless trilinear term
$\lambda H_u H_d S$ in the superpotential. When the scalar component
of the singlet superfield $S$ obtains a vacuum expectation value, a
bilinear term $\lambda H_u H_d <S>$ involving the two Higgs doublets
is naturally generated.  Furthermore, when this scalar component of
the chiral singlet superfield $S$ acquires a vacuum expectation value
of the order of the $SU(2)_L \times U(1)_Y$ breaking scale, it gives
rise to an effective value of $\mu$, $\mu_{eff} = \lambda <S>$, of the
order of electroweak scale.  However, the inclusion of the singlet
superfield leads to additional trilinear superpotential coupling
$(\kappa/ 3 ) S^3$in the model, the so called nonminimal, or
next-to-minimal~\cite{fayet, ellis89, drees89, nmssm1, nmssm2},
supersymmetric standard model~(NMSSM).  The absence of $H_u H_d$ term,
and the absence of tadpole and mass couplings, $S$ and $ S^2$ in the
NMSSM is made natural by postulating a suitable discrete symmetry.  
The NMSSM is attractive on account of the simple resolution it offers to the
$\mu$-problem, and of the scale invariance of its classical action in
the supersymmetric limit. Since no dimensional supersymmetric
parameters are present in the superpotential of NMSSM, it is the
simplest supersymmetric extension of the Standard Model in which the
electroweak scale originates from the supersymmetry breaking scale
only.  Its enlarged Higgs sector may help in relaxing the fine-tuning
and little hierarchy problems of the MSSM~\cite{nmssmft}, thereby
opening new perspectives for the Higgs boson searches at high energy
colliders~\cite{dob00, ellwang04, choi04, moort05}, and for dark
matter searches~\cite{gunion05}.

Since supersymmetry requires the introduction of superpartners of all
known particles in the SM, which transform in an identical manner
under the gauge group, there are additional Yukawa couplings in
supersymmetric models which violate~\cite{weinberg} baryon number~(B)
or lepton number~(L).  In the minimal supersymmetric standard model
there are both bilinear and trilinear lepton number violating Yukawa
terms in the superpotential. There are also trilinear baryon number
violating Yukawa terms in the superpotential.  All these terms are
allowed by renormalizability, gauge invariance and supersymmetry. In
MSSM, a discrete symmetry~\cite{farrar} called $R$-parity~($R_p$) is
invoked to eliminate these $B$ and $L$ violating Yukawa
couplings. However, the assumption of $R$-parity conservation at the
level of low energy supersymmetry appears to be {\it ad hoc}, since it
is not required for the internal consistency of supersymmetric models.

If we do not postulate $R_p$ conservation, then there are baryon and
lepton number number violating terms in the superpotential of NMSSM as
well.  What is perhaps interesting is the presence of an additional
lepton number violating trilinear superpotential
coupling~\cite{nmssmrp1, nmssmrp2} in this model which has no analog
in the MSSM with baryon and lepton number violation. It is, therefore,
important to study the implications of this additional lepton number
violating trilinear interaction term in the superpotential of NMSSM,
contrast the situation with MSSM with lepton number violation, and pin
down the possible differences with its predictions.

One of the far reaching implications of the lepton number violating
couplings in NMSSM concerns the physics of light neutrino states.  In
identifying the dominant contributions to the neutrino masses, and
suppression mechanisms, one must compare with the situation that
obtains in MSSM with bilinear lepton number violation. In NMSSM, the
three light neutrinos mix with $SU(2)_L \times U(1)_Y$ nonsinglet
gaugino and Higgsino fields as well as the gauge singlet fermionic
component of $S$, the singlino~($\tilde S$).  The resulting $8 \times
8$ mass matrix of the neutrino-gaugino-Higgsino-singlino has a see-saw
structure, which leads to a separable rank one effective mass matrix
for the light neutrinos, implying the presence of a single massive
Majorana neutrino. At one-loop order, there occur two main mechanisms
for generating masses for the Majorana neutrinos. One of these
involves only the matter interactions. The second mechanism involves
matter interactions in combination with the gauge interactions and
propagation of neutralinos and mixed sneutrino-Higgs boson system,
whose contribution depends sensitively on the soft supersymmetry
breaking couplings.  While both these mechanisms have the ability to
generate masses for the Majorana neutrinos, the latter one, initially
proposed in the context of MSSM by Grossman and
Haber~\cite{grosshaber99}, is expected to dominate.  In the case of
MSSM with bilinear lepton number~(or $R_p$) violation, the tree and
one-loop contributions to the neutrino masses, and their ability to
reproduce the experimental observations, have been extensively
discussed in the literature
~\cite{gross99,haberbasis,davidson00,abada02,
borzum02,chun02,grossm04, valle03, moha04}.

In this paper we carry out a detailed investigation of the nonminimal
supersymmetric standard model with lepton number violation. Since the
NMSSM has a unique lepton number violating trilinear coupling term in
the superpotential, one of the issues we want to address concerns the
implications of the neutrino masses and mixing for the NMSSM with
such a lepton number violating term.  Our purpose is to pin down the
features specific to this version of the  NMSSM and extract
constraints implied by the comparison with experimental data.  We
compare and contrast the situation in NMSSM with that of MSSM with
bilinear $R$-parity violation~(RPV)~\cite{grossm04}.  
Despite the presence of the singlino,
and its mixing with the neutrinos, a light mass Majorana neutrino
appears at the tree level. This is due to the constrained nature of
the couplings in the model. Nevertheless, as in the MSSM with lepton
number violation, one-loop contributions play an important role in
determining the neutrino mass spectrum.  The ability to reproduce
experimental observations is expected to set useful constraints on the
Higgs sector parameters of the NMSSM.  The situation differs from the
one that arises in the see-saw mechanism or the bilinear lepton number
violation in MSSM in that no dimensional mass parameters~(large or
small) are introduced. The neutrino Majorana masses arise from
dimensionless Yukawa couplings.  However, despite the presence of a
gauge singlet fermion that could play the r\^ole of a sterile
neutrino, whether an ultra light singlino mode, compatible with the
the cosmological bound on the summed mass of light neutrinos, $ \sum
_\nu m_\nu < 10 $ eV, does indeed occur is at variance with the
physical constraints on the NMSSM which rule out the possibility that
the lightest mode in the massive neutralino sector lies below $ O(50)$
GeV.  Thus, the understanding of neutrino physics provided by the
NMSSM with lepton number violation contrasts with that proposed in
models using the compactification moduli superfields~\cite{benak97} or
axion superfields~\cite{chun99} coupled gravitationally to the
observable sector modes.

\bigskip

This paper is organized as follows.  We begin in
Section~\ref{secmodel} with a discussion of the general structure of
the superpotential and soft supersymmetry breaking interactions in the
nonminimal supersymmetric standard model with lepton number
violation. In this Section we also discuss the local $Z_N$ cyclic
symmetries which can protect the NMSSM against $ B $ or $L$, or
combined $ B$ and $L$, number violating superpotential couplings
(Subsection~\ref{secdiss}).  We further elaborate on the general
approach to analyze the gauged cyclic group symmetries in the
Appendix~\ref{appex1} to the paper.  In Section~\ref{secneut} we
derive the tree level light neutrino mass spectrum that arises in this
model~(subsection~\ref{sectree}), and then obtain the one-loop
radiative corrections to the mass spectrum in
subsection~\ref{secloop}.  In Section~\ref{sechoriz} we present a
general discussion of the predictions from this model, which are based
on the consideration of Abelian horizontal symmetries for the flavor
structure of effective couplings.  Finally in Section~\ref{conclu} we
summarize our results and conclusions.

\section{NMSSM with baryon and lepton number violation}
\label{secmodel} 

\subsection{The superpotential}
\label{secpot} 
In this section we recall the basic features of NMSSM with baryon and
lepton number violation, and establish our notations and conventions.
The superpotential of the model is written as
\begin{equation}\label{nmssmw}
W_{NMSSM} = (h_U)_{ab} Q^a_L \overline{U}^b_R H_u + (h_D)_{ab} Q^a_L
\overline{D}^b_R H_d + (h_E)_{ab} L^a_L \overline{E}^b_R H_d + \lambda
S H_d H_u - \frac{\kappa}{3}S^3,
\end{equation}
where $L,\, Q, \, \overline{E},\, \overline{D},\, \overline{U}$ denote
the lepton and quark doublets, and anti-lepton singlet, d-type
anti-quark singlet and u-type anti-quark singlet, respectively.  In
Eq.~(\ref{nmssmw}), $(h_U)_{ab}$, $(h_D)_{ab}$ and $(h_E)_{ab}$ are
the Yukawa coupling matrices, with $a,\, b,\, c$ as the generation
indices.  Gauge invariance, supersymmetry and renormalizability allow
the addition of the following $L$ and $B$ violating terms to the
superpotential (\ref{nmssmw}):
\begin{eqnarray}
W_L &=& \tilde\lambda_{a} L_a H_u S + {1\over 2}\lambda_{abc} L^a_L
L^b_L \overline{E}^c_R + \lambda'_{abc} L^a_L Q^b_L\overline{D}^c_R,
\label{Lviolating} \\ W_B &=& {1\over 2}\lambda''_{abc}
\overline{D}^a_R \overline{D}^b_R \overline{U}^c_R, \label{Bviolating}
\end{eqnarray}
where the notation~\cite{nmssmrp1, nmssmrp2} is standard.  We note
that there is an additional $L$-violating term with the dimensionless
Yukawa coupling $\tilde\lambda_a$ in (\ref{Lviolating}) which does not
have an analogue in the MSSM.  This term can be rotated away into the
R-parity conserving term $\lambda S H_uH_d$ via an $SU(4)$ rotation
between the superfields $H_d$ and $L_a$. However, this rotation must
be performed at some energy scale, and the term is regenerated through
the renormalization group equations.  The Yukawa couplings
$\lambda_{abc}$ and $\lambda''_{abc}$ are antisymmetric in their first
two indices due to $SU(2)_L$ and $SU(3)_C$ group symmetries,
respectively.
     
The supersymmetric part of the Lagrangian of NMSSM with baryon and
lepton number violation can be obtained from the superpotential
(\ref{nmssmw}), (\ref{Lviolating}) and (\ref{Bviolating}) by the
standard procedure. In addition to this supersymmetric Lagrangian,
there are soft supersymmetry breaking terms which include soft masses
for all scalars, gaugino masses, and trilinear scalar couplings,
respectively.  These can be written as \begin{eqnarray}
{V}_{\rm{soft}} = - \cal{L}_{\rm{soft}} &=& \left[M_Q^{ab2}
\tilde{Q}^{a*}\tilde{Q}^b_L + M_U^{ab2} \tilde{\overline{U}}^a_R
\tilde{\overline{U}}^{b*}_R + M_D^{ab2} \tilde{\overline{D}}^a_R
\tilde{\overline{D}}^{b*}_R + M_L^{ab2} \tilde{L}^{a*}_L \tilde{L}^b_L
+ M_E^{ab2} \tilde{\overline{E}}^a_R \tilde{\overline{E}}^{b*}_R
\right.  \nonumber \\ && + \left. m^2_{H_d} H_d^* H_d + m^2_{H_u}
H_u^* H_u + m^2_S S^* S \right] \nonumber \\ & + & \left[\frac{1}{2}
M_s \lambda^s \lambda^s + \frac{1}{2} M_2 \lambda^w \lambda^w +
\frac{1}{2} M_1 \lambda' \lambda' \right]\nonumber \\ & +
&\left[(A_U)_{ab}(h_U)_{ab} \tilde{Q}^a_L \tilde{\overline{U}}^b_R H_u
+(A_D)_{ab}(h_D)_{ab} \tilde{Q}^a_L \tilde{\overline{D}}^b_R H_d
\right.  \nonumber \\ && + \left. (A_E)_{ab}(h_E)_{ab} \tilde{L}^a_L
\tilde{\overline{E}}^b_R H_d - A_{\lambda} \lambda S H_d H_u -
\frac{A_{\kappa}}{3} {\kappa} S^3 \right] \nonumber \\ & & + \left[ -
(A_{\tilde\lambda _{a}}) \tilde\lambda_a \tilde{L}^a_L H_u S + {1\over
2}(A_\lambda)_{abc}\lambda_{abc} \tilde{L}^a_L \tilde{L}^b_L
\tilde{\overline{E}}^c_R + (A_{\lambda'})_{abc}\lambda'_{abc}
\tilde{L}^a_L \tilde{Q}^b_L \tilde{\overline{D}}^c_R \right] \nonumber
\\ &+& \left[ {1\over 2}(A_{\lambda''})_{abc}\lambda''_{abc}
\tilde{\overline{D}}^a_R \tilde{\overline{D}}^b_R
\tilde{\overline{U}}^c_R \right] + \ H.\ c. ,
\label{soft}
\end{eqnarray}
where a tilde over a matter chiral superfield denotes its scalar
component, and the notation for the scalar component of the Higgs
superfield is the same as that of the corresponding superfield.  We
note that the soft supersymmetry breaking gaugino masses have been
denoted by $M_1, M_2,$ and $M_s$ corresponding to the gauge groups
$U(1)_Y, \, SU(2)_L, \,$ and $SU(3)_C$, respectively.  We have chosen
the sign conventions for the soft trilinear couplings involving the
gauge singlet field in Eq.(\ref{soft}) which are different from those
used in Ref~\cite{nmssmrp2}.
                                                                                
The dimension-4 terms in the superpotentials (\ref{Lviolating}) and
(\ref{Bviolating}) are the most dangerous terms for nucleon decay, and
some of them must be suppressed. This leads to
constraints~\cite{chemtob} on the different couplings $\lambda_{abc},
\lambda'_{abc}$, and $\lambda''_{abc}$, but considerable freedom
remains for the various $B$ and $L$ violating couplings.  Furthermore,
there are dimension-5 operators~\cite{nmssmrp2} which may lead to
nucleon decay suppressed by $1/M$, where $M$ is some large mass scale
at which the $B$ and $L$ violation beyond that of NMSSM (and MSSM)
comes into play.  Some of these dimension-5 operators may also lead to
unacceptable nucleon decay if their coefficients are of order unity,
and therefore must be suppressed.  We shall not consider here the
dimension-5 operators, but instead concentrate on the dimension-4
lepton number violating terms (\ref{Lviolating}) only.

As noted above, there is an additional trilinear $L$-violating term
with the dimensionless Yukawa coupling $\tilde\lambda_a$ in
(\ref{Lviolating}) which does not have an analogue in the MSSM.  This
term can be rotated away into the R-parity conserving term $\lambda S
H_uH_d$ via an $SU(4)$ rotation between the superfields $H_d$ and
$L_a$. However, this rotation must be performed at some energy scale,
and the term is regenerated through the renormalization group
equation~\cite{nmssmrp1, nmssmrp2} \bea
\frac{d\tilde\lambda_3}{d\ln\mu}&=&\frac{1}{16\pi^2}
\left[\left(3h_t^2+h_{\tau}^2
+4\lambda^2+2{\kappa}^2+4\tilde\lambda_3^2+\lambda_{233}^2+3\lambda_{333}^{'2}
\right)\tilde\lambda_3\right. \nonumber\\
&&\left.+3h_b\lambda\lambda_{333}^{'} -\left(3g_2^2+
\frac{3}{5}g_1^2\right)\tilde\lambda_3\right],
\label{RGL}
\eea where $(h_U)_{33} \sim h_t, \, (h_D)_{33} \sim h_b, \, (h_L)_{33}
\sim h_{\tau}$. Here $g_2, \, g_1 \, $ are the $SU(2)_L$ and $U(1)_Y$
gauge couplings, and the meaning of other quantities is obvious. For
simplicity, we have retained only the highest generation trilinear
couplings in (\ref{RGL}).

It is important to point out two distinctive features of the present
model relative to the MSSM. First, in NMSSM no distinction is made
between the bilinear and trilinear lepton number violation, since the
bilinear Lagrangian terms, $\mu H_d H_u + \mu_i L_i H_u , \ [\mu = \l
<S>, \ \mu_i= \tilde \l_i <S>] $, can arise as effective couplings
once the singlet scalar field component of $S$ acquires a finite VEV.
The wide hierarchies between the lepton number conserving and
violating couplings, of the expected size $ \tilde \l _i /\l = \mu _i /\mu
\sim 10^{-6} $, arise from the hierarchies of the dimensionless Yukawa
couplings. Second, the lepton number violating trilinear operator $L_i
H_u S$ has the ability to radiatively induce other trilinear lepton
number violating couplings, which is precluded for the bilinear
operator $L_i H_u$.  This property may be used to justify a scheme
where naturally suppressed trilinear couplings $\l _{ijk}, \ \l '
_{ijk} $ occur as a result of being set to zero at some large mass
scale~(gauge unification scale), and receive small finite radiative
corrections from the gauge singlet coupling $\tilde \l _i L_i H_d S$.
This possibility can be established on a quantitative basis by
examining the one-loop renormalization group equations for the
trilinear coupling with maximal number of third generation
indices~\cite{nmssmrp2} \bea (4\pi )^2 {\dh \l _{233} \over \dh \log
Q} &=& \l _{233} [ 4 h_\tau ^2 +4 \l _{233} ^2 + 3 \l _{333} ^{'2}
+\tilde \l ^2_3 -( {9\over 5} g_1 ^2+ 3 g_2^2 ) ], \nonumber \\ (4\pi
)^2 {\dh \l '_{333} \over \dh \log Q} & = & \l '_{333} [h_t ^2 + 6h_b
^2 +h_\tau ^2 + \tilde \l ^2_3 +\l _{233} ^2 + 6 \l _{333} ^{'2}
\nonumber \\ && - ( {16 \over 3} g_3^2 + 3 g_2^2 + {7\over 15} g_1 ^2
) ] + h_b \l \tilde \l _3 .  \eea Looking for solutions of these
equations with suitably large values of the trilinear couplings at
some large mass scale, for instance, gauge coupling unification scale,
might reveal the presence of infrared fixed points which would then
serve as upper bounds on the weak scale values of these couplings.

To establish the lepton number violating nonminimal supersymmetric
standard model on a firmer basis, it is important to determine whether
there are  discrete symmetries respecting the postulated interaction
superpotential which can be regarded as   local or gauged
symmetries~\cite{wilczek} obeying the anomaly cancellation conditions.
As is known, the gauged discrete symmetries enjoy a natural protection
against breaking by non-perturbative quantum effects initiated by the
gravitational interactions, and against the emergence of massless
Nambu-Goldstone bosons from the spontaneous symmetry breaking.
Another advantage lies in evading the cosmological domain wall problem
by the removal of classical domain wall solutions as a result of the
gauge equivalence of degenerate vacua.  We recall that if stable
domain wall solutions were present, the production of cosmic domain
walls in the early universe would result in a contribution to the
present day mass density of the universe which exceeds the critical
density.  The case of Green-Schwarz (GS) anomalous gauged discrete
symmetries is special in that although domain wall solutions do exist
in gauge field theories satisfying such global type symmetries, the
instanton tunneling effects present in these theories lift the
degeneracy of vacua so as to render the solutions
unstable~\cite{presk91,banksdine91}.  The above resolution applies
independently of the familiar one invoking the domain wall dilution
during a cosmic inflation era.  We also note that the discrete gauge
symmetries have been used in connection with various naturalness
problems, such as the doublet-triplet splitting in unified gauge
models, the stabilization of axion symmetries, or the construction
of flavor symmetries realized via Froggatt-Nielsen
mechanism~\cite{castano,dreiner}.  Following the work by Ib\'a\~nez and
Ross~\cite{iban92,iban95}, we are led here to consider the so-called
generalized parities (GPs) of the NMSSM which forbid part or all of
the dangerous couplings at the renormalizable level~\cite{lola93}. In
the following Subsection we shall describe the construction of the
generalized parities for the NMSSM in detail.

\subsection{Discrete symmetries}
\label{secdiss} 

In this Subsection we consider the $Z_N$ cyclic local symmetries which
can protect the NMSSM against $B$, or $L$, or combined $B$ and $L$
number violating superpotential couplings.  Demanding consistency with
the anomaly cancellation conditions sets highly non-trivial
constraints on the generalized baryon, lepton and matter parity
symmetries (designated as GBP, GLP and GMP) for the appropriate
superpotential.  These are examined by making use of the approach of
Ib\'a\~nez and Ross~\cite{iban92,iban95}, which is developed in
Appendix~\ref{appex1} for the NMSSM.  Before presenting our results, 
we shall outline the problem in general and introduce our notations.

Besides the regular R-parity conserving~(RPC) trilinear matter-Higgs
Yukawa couplings, $ Q H_uU^c,\ Q H_d D^c,\ LH_d E^c$, and the
dangerous R-parity violating (RPV), and $B$ and $L$ violating,
couplings $U^cD^cD^c$ and $ LLE^c ,\ LQD^c $, the renormalizable
superpotential of the NMSSM includes the trilinear couplings $H_dH_u S
$ and $L H_u S $, but excludes the dimensional superpotential
couplings $ S,\ S^2$ and $H_d H_u $.  Thus, in addition to the weak
hypercharge $Y$, the regular couplings conserve three $U(1)$ charges.
A convenient basis for the generators of the corresponding 
3-dimensional vector space is given by the $U(1)$ charges 
$\hat R ,\ \hat A,\ \hat L$, which are defined in the table 
of Appendix~\ref{appex1}.  The cyclic
$Z_N$ group elements $ R = e ^{i\a _R \hat R } ,\ A = e ^{i\a _A \hat
A },$ and $L = e ^{i\a _L \hat L }$ are defined by restricting the
complex phase rotation angles to the values $ \a _R= 2\pi m /N ,\ \a
_A= 2\pi n /N ,\ \a _L= 2\pi p /N,$ with $m,n,p $ being integers.  The
generators of the independent $Z_N$ multiplicative symmetry groups are
thus of the form, $ g = R^m A^n L^p = g_{PQ} ^n R^{m-2n} L^p ,\ [
g_{PQ}= R^2 A]$ with specific (modulo $N$) relations linking the three
integers $m, n, p$, depending on the allowed subset of dangerous
couplings.  For the ordinary symmetries~(${\bf O}$), the charges are
given explicitly by $ \hat g (Q)= 0,\ \hat g (U^c)= -m,\ \hat g (D^c)=
m-n,\ \hat g (L)= -n-p,\ \hat g (E^c)= m+p,\ \hat g (H_d)= -m+n,\ \hat
g (H_u)= m,\ \hat g (S)= mx +ny+pz$. The charges $ x, y, z$ assigned
to the gauge singlet $S$ must obey the selection rules, $ x+y+z \ne
0,\ 2(x+y+z ) \ne 0,\ 3( x+y+z ) = 0 ,\ x+y+z+n = 0 $.  A special
r\^ole is played by the Peccei-Quinn like charge, $(PQ)= 2 \hat R
+\hat A$, which has a finite color group anomaly and is conserved by
all the NMSSM couplings except $ S^3 $, so a massless axion mode would
arise only if the cubic coupling were absent, $\kappa =0$.  The
general form of the GBP, GLP and GMP generators is displayed in
eq.~(\ref{eqgps2}) of Appendix~\ref{appex1}.  The soft supersymmetry
breaking terms, which are generated via couplings to the goldstino
spurion superfield $X$ of form, $ V _{soft} \sim [X W]_F + H.\ c. =
F_X W + H.\ c. $, are automatically protected for ordinary symmetries
(${\bf O}$), and this protection remains valid for the R symmetries
(${\bf R}$) as well, provided one assigns the R charge $Q(X)=0$, and
hence $ Q(F_X) = 2$.
\begin{center} 
\begin{table}{htb} 
\caption{\it The solutions for the generalized flavor blind $Z_N$
parity symmetries of the NMSSM which cancel the mixed gauge anomalies
$ \cala _{3,2,1}$ only. We have made the choice $ N_g=3, \ N_{2h}=1$
for the number of quark and lepton generations and Higgs boson pairs,
respectively, and $k_1={5\over 3} $ for the hypercharge normalization.
The cancellation conditions for the gravitational and non-linear
anomalies, $\cala _{grav },\ \cala _{Z^2}, \ \cala _{Z^3}$ are not
obeyed in general.  The four rows give the solutions for the GBP, GLP,
and GMP generators for symmetries of four distinct types: ordinary
anomaly free $({\bf O})$, ordinary Green-Schwarz anomalous $({\bf
O/GS})$, $R$ symmetry anomaly free~$({\bf R})$, and anomalous $R$
symmetry~$({\bf R/GS})$, respectively.  The entries display integer
(modulo $N$) parameters $ (m,n,p) $ for the $Z_N$ generator, and in
the suffix we have given the values of the finite anomaly coefficients
$\cala _{grav}, \ \cala _{Z^2}$. For simplicity, we have not quoted
the generally finite chiral anomaly coefficient $\cala _{Z^3}$. For
the GMP symmetries, additional  solutions of same order $N$  as 
those quoted in the Table  below  also arise, as 
discussed in the text.}
\vskip 0.5 cm
\begin{tabular}{|c|c|c|c|c|}  
\hline &&&& \\ $ Z_N$ {\bf Type} & $\quad N \quad $ & {\bf GBP} & {\bf
GLP} & {\bf GMP} \\ &&&&\\ \hline &&&& \\ {\bf O } & 9 & $ ( 1, 3,
4)_{ -51,-87}$ &$ ( 6, 3, 4)_{ -36,-117}$ & $ ( 0, 3, 4)_{-54, -81} $
\\ && $ ( 5, 6, 2)_{ -75,-348}$ &$ ( 3, 6, 2)_{-81,-252}$ & $ ( 2, 3,
4)_{-48,-93} $ \\ && $ ( 2, 6, 8)_{-102,-348} $ & $ ( 3, 6,
8)_{-99,-360}$ & $ ( 3, 3, 4)_{-45,-99}$ \\ &&&& \\ \hline &&&& \\
{\bf O/GS} & 7 & $( 0, 6, 2)_{-126,-108}$ & $ ( 5, 6, 2)_{-141,-348}$
& $ ( 1, 6, 2)_{-129,-156},\ ( 2, 6, 2)_{-132,-240}$ \\ &&&& $ ( 3, 6,
2)_{-135,-252},\ ( 4, 6, 2)_{-138,-300} $ \\ &&&& \\ \hline &&&& \\
{\bf R} & 12 & $ ( 8, 6, 6)_{-131,-560}$ & $ ( 2, 6, 6)_{-149,-320}$ &
$ ( 0, 6, 6)_{-155,-240},\ ( 1, 6, 6)_{-152,-280}$ \\ &&&& \\ \hline
&&&& \\ {\bf R/GS} & 11 & $ ( 5, 6, 2)_{-160,-368}$ & $ ( 3, 6,
2)_{-154,-240}$ & $ ( 0, 6, 2)_{-145,-48},\ ( 1, 6, 2)_{-148,-112}$ \\
&&&& \\ \hline
\end{tabular}
\label{tablegps}
\end{table} 
\end{center} 
\vskip 0.5 cm

Having classified the Abelian GP generators of fixed order $N$ in
terms of the three integers $ (m, n, p) $, our next task is to
select the solutions satisfying the quantum anomaly conditions.
These  are expressed in terms of an over-determined system
of linear and nonlinear equations for $ m, n, p $.  We have
developed a numerical program to solve the anomaly
cancellation conditions for the generalized parity generators (GBP,
GLP, GMP) of four types.  Unless stated otherwise, the search has been
restricted to the case involving three quark and lepton generations,
$N_g=3$, and a single pair of Higgs boson doublets, $ N_{2h} =1$.
While our presentation of results will be limited to cyclic groups of
order $ N\le 15$, we note that higher order solutions occur at integer
multiples of the low order solutions, and as such they do not reveal
novel features.

If we demand that all the anomaly constraints are satisfied, then we
find that, in general, no solutions exist, even if one is willing to
push the search to high enough group orders, say $N \leq 30$.
However, one must realize that the various anomaly cancellation
conditions need not all be placed on the same footing.  The linear
anomalies $ \cala _3,\ \cala _2,\ \cala _1,\ \cala _{grav}$ have an
obvious priority over the others, to the extent that these identify
with the selection rules obeyed by the determinant interactions of
fermions mediated by the classical instanton
solutions~\cite{presk91,banksdine91} of the non-Abelian gauge theory
factors.  More importantly, these conditions are independent of the
$Z_N$ charge normalization, in contrast to the less physically
motivated non-linear anomalies, $ \cala _{Z^2} $ and $ \cala _{Z^3} $,
which thus depend on the spectrum of massive decoupled modes.  Among
the linear anomalies, the gravitational and Abelian gauge $U(1)_Y$
anomalies, $ \cala _{grav} $ and $ \cala _1 $, are believed to be less
robust than the non-Abelian ones, $ \cala _3,\ \cala _2 $.  Indeed,
$\cala _1$ is sensitive to the normalization of the hypercharge which
remains a free parameter as long as one is not concerned with the
gauge group unification.  Also, $ \cala _{grav},\ \cala _{Z^3}$ are
sensitive to contributions from additional gauge singlet fields, which
could either belong to the observable sector, such as the singlet $S$,
or to the hidden sector, and hence coupled only through the
gravitational interactions.  While the non-linear anomalies $ \cala
_{Z^2} ,\ \cala _{Z^3}$ are both sensitive to the ambiguity which
arises due to the arbitrary $Z_N$ charge normalization, $ \cala
_{Z^3}$ is also sensitive to presence of gauge singlets.

We now discuss our results. These are displayed in
Table~\ref{tablegps} for the four different realizations of the three
generalized parity symmetries in terms of the generator indices $
(m,n,p) $.  Let us start first with the ordinary symmetries {\bf O}.
While it proves impossible to solve the complete set of equations, as
already indicated, solutions do arise in very large numbers if one
chooses to cancel the non-Abelian gauge anomalies $ \cala _{3} ,\
\cala _{2} $ only.  At this point, we mention that in the case of
ordinary symmetries, and only for this case, the cubic anomaly $ \cala
_{Z^3}$ automatically cancels once the equations for $ \cala _{3} ,\
\cala _{2} $ are satisfied.  The option of cancelling only the mixed
gauge anomalies, $\cala _{3},\ \cala _{2},\ \cala _{1}$,~(including
the $ \cala _{Z^3} $ for the {\bf O} symmetries is much more
restrictive, but still yields solutions.  The first realization of
{\bf O} symmetry occurs at the group order $ N=9$ with $3$ GBP and GLP
solutions and some $22$ GMP solutions.  We have quoted in
Table~\ref{tablegps} the values for the gravitational anomalies and
$\cala _{Z^2}$ which remain generally uncanceled.  For instance, the
GBP generator $ (m,n,p)= (1,3,4) \ \text{mod} \ (9)$, has $\cala
_{grav}= -81 =0 \ \text{mod}\ (9 ) $, and hence a single uncanceled
anomaly, $ \cala _{Z^2} = -87 = -6 \ \text{mod} \ (9 ) $.  At higher
group orders, a restricted number of similar {\bf O} solutions arise
at orders $N=18 $ and $N=27$.  For the GMP case we have displayed in
the table only a subset of the solutions. In fact, the {\bf O} GMP
solutions for the $Z_9$ group can be grouped into the three families
of generators, $ (m,3,4), \ [m=0,2,3,4,5,7] ;\ (m,6,2), \
[m=0,1,2,4,6,7] ;\ (m,6,8), \ [m=0,1,4,5,6,7] $.

Having a fixed $N$ solution is not enough, since we must
still solve the equations for the $S$ field charges.  A large number
of solutions exist, in general, for the {\bf O} symmetries under
discussion.  For instance, with the GBP solution $ (m,n,p)= (1,3,4) \
\text{mod} (9)$, the equations $ x +3y + 4 z=-3 \ \text{mod} (9) , 3 (
x +3y + 4 z)=0\ \text{mod} \ 9 $ admit about $80$ different solutions
of which we quote an illustrative sample: $(x,y,z)= (0,0,6),\
(0,1,3),\ (0,2,0), (0,3,6), \ (0,4,3) $.  The two other solutions,
namely $ (m,n,p)= (5,6,2), \ \ (2,6,8) \ \text{mod} \ 9 $, have
similar features.

The case of Green-Schwarz anomalous symmetries {\bf O/GS} is more
constrained than the non-anomalous case discussed above.  Unless one
excludes the model dependent anomalies $\cala _{Z^2}, \ \cala _{Z^3}$
and $ \cala _{grav} $, no solutions exist.  However, by restricting
again to the mixed gauge anomaly cancellation conditions only, which
we express in terms of vanishing $N$ linear combinations modulo $N$, $
\cala _3 / k_3 - \cala _2 / k_2 ,\ k_1 \cala _3 / k_3 - \cala _1,\ k_1
\cala _2 / k_2 - \cala _1$, and setting the normalization parameters
for the SM gauge group factors at the rational values, $ k_3=k_2=1,\
k_1={5\over 3}$, we find a single GBP solution appearing first for the
group $Z_7$ with the generator $(m,n,p)= (0, 6, 2) $. While this
generator turns out to have a vanishing gravitational anomaly,
$k_{grav} \cala _3 / k_3 -\cala _{grav} = - 126 =0 \ \text{mod} \ 7$,
it still exhibits an uncanceled non-linear anomaly, $ \cala _{Z^2}
=-108 =-3 \ \text{mod} \ 7 $.  Note that the next group order at which
solutions appear is the integer multiple of the above with $ N=14 $.
Unfortunately, proceeding to the next stage of solving for the $S$
field charges, we find it impossible to solve the relevant equations
for the integers $x, y, z$.  It is possible that this feature may be
cured by adding an extra gauge singlet field.

The cyclic R parity discrete symmetries, {$\bf R$} and {$\bf R/GS$},
are more severely constrained than the ordinary ones.  As seen on
Table~\ref{tablegps}, unique solutions are found at orders $12$ and
$11$, respectively, if one chooses to cancel the gauge anomalies only,
while leaving $\cala _{grav}$ and the non-linear anomalies
uncancelled.  The {$\bf R$} GMP solutions for the group $Z_{12}$
include the family $(m,6,6), \ [m=0,1,3,4,5,7,7,9,10] $, and the {$\bf
R/GS$} GMP solutions for the group $Z_{11}$ include the family
$(m,6,2), \ [m=0,1,2,4,6,7,8,9] $.  For both anomaly free and
anomalous cases, we fail to find solutions for the equations for the
$S$ field charges $x,y,z$.

In closing the discussion of results, we note that practically all the
GMP solutions in Table~\ref{tablegps} forbid the dimension $5$
dangerous operators $QQQL$ and $U^cD^cU^cE^c$.  This is easily
established by noting the corresponding selection rules, which require
for ${\bf O}$ and ${\bf R}$ symmetries vanishing values for the total
charges, $ \hat g(QQQL)= -(n+p),\ \hat g( U^cD^cU^cE^c)= -(n-p) $ and
$ \D (QQQL)= - (2+n+p),\ \D ( U^cD^cU^cE^c)= - (2+n-p) $, where $ \D (
\Phi ^M ) = \hat {\tilde g } (W) - (2-M) $.

It is of interest to find out whether, by slightly modifying our
search strategy, alternative options could exist.  One might first
consider solving the anomaly cancellation equations by setting the
number of generations at the smaller values, $ N_g =1$ or $N_g
=2$. These solutions can be combined into quark and lepton generation
dependent direct products, $ (Z_N)_{N_g=1} ^3 $ or $(Z_N)_{N_g=2}
\times (Z_N)_{N_g=1}$.  This option is not promising, however, since
the solutions for $ N_g < 2 $ are even more scarce than for $ N_g =3
$.  Thus, for GBP with $N_g=2$, the first solution for ${\bf O} $
symmetries occurs at $ N=16$, with $ (m,n,p)= (14,12,6)$, for ${\bf R}
$ symmetries at $ N=12$, with $ (m,n,p)= (8,6,6)$, and for ${\bf
R/GS}$ symmetries at $ N=5$ with $ (m,n,p)= (2,4,2)$.  The case with
$N_g=1$ does not have any solutions.

Another option consists in enforcing the modulo $N$ cancellation of
$\cala _1$ by adjusting the hypercharge normalization. As already
noted, the freedom gained in relaxing $\cala _1$ anomaly cancellation
constraint vastly increases the space of solutions.  Changing the
hypercharge normalization $Y\to c Y $ induces the modifications $\cala
_1 \to c ^2 \cala _1 \ \Longrightarrow \ k_1\to k'_1 =c^ 2k_1$.
Specifically, given an {\bf O} charge generator $\hat g$ with an
uncanceled hypercharge anomaly, $\cala _1 \ne 0 \ \text{mod} (N) $,
we can salvage the situation by transforming $Y \to cY $ so that $ c^2
\cala _1 =0 \ \text{mod} (N) $. For the anomalous {\bf GS} symmetry,
the same reasoning applies to the linear combination $ c^2 \cala _1 -2
k '_1 =0 \ \text{mod} \ N $.  Of course, within a non-minimal grand
unification or a string theory model in which some freedom is left for
the hypercharge normalization,
one must still consider how well the asymptotic prediction for the
weak angle parameter, $ \sin ^2 \t _W \simeq k_2 /(k_2 + k '_1) $,
fits in with the observed value.  Considering, for concreteness, the
illustrative case where $ \cala _1 = -pN +\nu , \ [p, \nu \in Z ^+ ] $
and changing $ Y\to c Y $, so that $ \cala ' _1 = c^2 \cala _1 = -pN
$, fixes the rescaling factor as $c^2 = {k_1'\over k_1} = {-pN \over
-pN +\nu }\simeq 1 + {\nu \over pN}$, and hence the modified
asymptotic value of the weak angle as $ \sin^2 \t _W \simeq {3/8 \over
1 + {5 \nu \over 8 pN }} $, where we have assumed, for simplicity, $
\nu << N$.

The case of an uncanceled $\cala _{grav} $ anomaly may be treated by
adding observable or hidden sector chiral supermultiplet singlets, as
already hinted above.  Both kinds of singlets affect only $ \cala
_{grav} , \ \cala _{Z^3}.$ Thus, given some generator with uncanceled 
gravitational and chiral anomalies, $\cala _{grav} \ne 0 \ \text{mod}
\ N ,\ \cala _{Z^3}\ne 0 \ \text{mod} \ N $, one can attempt to rescue
this solution by including an extra hidden sector singlet $S_1$ with
$\hat R,\hat A,\hat L$ charges $ x_1, y_1, z_1$, and solving the
equations, $\cala _{grav} + \cals _1 =0 \ \text{mod} \ N , \ \cala
_{Z^3} + \cals ^3 _1 =0 \ \text{mod} \ N $ with $\cals = mx+ny+pz $.
Obvious modifications hold for the Green-Schwarz (GS) anomalous
symmetries and for $R$ symmetries.

The option of extending the matter field content of the low energy
theory by adding extra vector multiplets is also very efficient in
relaxing the anomaly constraints.  Indeed, we recall that the string
theory models achieve consistency thanks to the presence of extra
charged or singlet modes in the massless particle spectrum.  To
conclude, we note that interesting generalizations of our discussion
would be to consider direct product of cyclic groups, $Z_N \times
Z_M$, or lepton flavor dependent cyclic groups.

\section{Phenomenological implications and neutrino masses} 
\label{secneut} 
\subsection{\bf Tree level neutrino masses }
\label{sectree} 
After spontaneous breaking of the $SU(2)_L \times U(1)_Y$ gauge
symmetry via the vacuum expectation values of the scalar components of
$H_u, H_d,$ and $S$, the gauginos and Higgsinos mix with neutrinos.
The resulting lepton number violating neutrino-gaugino-Higgsino mass
matrix receives contributions from gauge interactions and the
superpotential
\begin{equation}
\label{massmatrix}
W_{\nu} = \lambda S H_d H_u + \tilde\lambda_{a} L_a H_u S -
\frac{\kappa}{3}S^3,
\end{equation}
which arises from the last two terms of (\ref{nmssmw}) and the first
term of (\ref{Lviolating}), respectively. The resulting mass terms of
the neutrino-gaugino-Higgsino system can be written as \bea {\cal
L}_{mass} & = & [ \lambda x \tilde H_u \tilde H_d + \lambda v_u \tilde
H_d \tilde S + \lambda v_d \tilde H_u \tilde S - \kappa x \tilde S
\tilde S + H. c.] \nonumber \\ & + & \frac{i g_2 \lambda^3}{\sqrt 2}
[v_d \tilde H_d - v_u \tilde H_u +H.c.]  - \frac{i g_1 \lambda'}{\sqrt
2} [v_d \tilde H_d - v_u \tilde H_u +H.c.]  \nonumber \\ & + & \sum_a
\tilde \lambda_a [ x \nu_a \tilde H_u + v_u \nu_a \tilde S + v_a
\tilde H_u \tilde S + H.c.] \nonumber \\ & + & \frac{i g_2
\lambda^3}{\sqrt 2} [\sum_a v_a \nu_a + H.c.]  - \frac{i g_1
\lambda'}{\sqrt 2} [\sum_a v_a \nu_a + H.c.],
\label{lagrangian}
\eea where $\lambda^3$ is the third component of the $SU(2)_L$ gaugino
$\lambda^w$, and $\lambda'$ is the $U(1)$ gaugino. In
(\ref{lagrangian}) we have used the following notation \bea v_u & = &
<H_u^0>, ~~~ v_d = <H_d^0>, ~~~ \tan\beta = \frac {v_u}{v_d},
\nonumber \\ x &=& <S>, \ v_a = <\tilde\nu_a>,
\label{defs1}
\eea while rest of the symbols have their usual meaning. Using
(\ref{lagrangian}), we find the resulting $8\times 8$
neutralino-neutrino mass matrix in the field basis $(-i \lambda', -i
\lambda^3, \tilde H_u, \tilde S, \tilde H_d, \nu _e, \nu _{\mu}, \nu
_{\tau}) $ as \bea {\bf M_N} & = &\left[
\begin{array}{cccccccc}
M_1 & 0 & {g_1 v_u\over \sqrt 2} & 0 & {-g_1 v_d\over \sqrt 2} & {-g_1
v_1\over \sqrt 2} & {-g_1 v_2\over \sqrt 2} & {-g_1 v_3\over \sqrt 2}
\\ 0 & M_2 & {- g_2 v_u\over \sqrt 2} & 0 & {g_2 v_d\over \sqrt 2} &
{g_2 v_1\over \sqrt 2} & {g_2 v_2\over \sqrt 2} & {g_2 v_3\over \sqrt
2}\\ {g_1 v_u\over \sqrt 2} & {-g_2 v_u\over \sqrt 2} & 0 & Y &
-\lambda x & -\tilde\lambda_1 x & -\tilde\lambda_2 x &
-\tilde\lambda_3 x \\ 0 & 0 & Y & 2\kappa x & -\lambda v_u &
-\tilde\lambda_1 v_u & -\tilde\lambda_2 v_u & -\tilde\lambda_3 v_u \\
{-g_1 v_d\over \sqrt 2} & {g_2 v_d\over \sqrt 2} & -\lambda x &
-\lambda v_u & 0 & 0 & 0 & 0 \\ {-g_1 v_1\over \sqrt 2} & {g_2
v_1\over \sqrt 2} & -\tilde\lambda_1 x & -\tilde\lambda _1 v_u & 0 & 0
&0 & 0\\ {-g_1 v_2 \over\sqrt 2} & {g_2 v_2\over \sqrt 2} &
-\tilde\lambda_2 x & -\tilde\lambda_2 v_u & 0 & 0 & 0 & 0 \\ {-g_1
v_3\over \sqrt 2} & {g_2 v_3\over \sqrt 2} & -\tilde\lambda_3 x &
-\tilde\lambda_3 v_u & 0 & 0 & 0 & 0
\end{array}
\right],
\label{nmatrix}
\eea where \bea Y & = &-\lambda v_d - \sum_a \tilde \lambda_a v_a.
\label{defs2}
\eea The mass matrix (\ref{nmatrix}) can be written in the form \bea
{\bf M_N} & = & \left[
\begin{array}{cc} 
\cal{M}_\chi & m^T \\ m & 0_{3\times 3}
\end{array}
\right],
\label{seesaw} 
\eea where
\bea \cal{M}_{\chi} & = &\left[
\begin{array}{ccccc}
M_1 & 0 & {g_1 v_u\over \sqrt 2} & 0 & {-g_1 v_d\over \sqrt 2}\\ 0 &
M_2 & {- g_2 v_u\over \sqrt 2} & 0 & {g_2 v_d\over \sqrt 2} \\ {g_1
v_u\over \sqrt 2} & {-g_2 v_u\over \sqrt 2} & 0 & Y & -\lambda x \\ 0
& 0 & Y & 2\kappa x & -\lambda v_u \\ {-g_1 v_d\over \sqrt 2} & {g_2
v_d\over \sqrt 2} & -\lambda x & -\lambda v_u & 0\\
\end{array}
\right],
\label{neutralino}
\eea and \bea m^T & = &\left[
\begin{array}{ccc}
{-g_1 v_1\over \sqrt 2} & {-g_1 v_2\over \sqrt 2} & {-g_1 v_3\over
\sqrt 2} \\ {g_2 v_1\over \sqrt 2} & {g_2 v_2\over \sqrt 2} & {g_2
v_3\over \sqrt 2}\\ -\tilde\lambda_1 x & -\tilde\lambda_2 x &
-\tilde\lambda_3 x \\ -\tilde\lambda_1 v_u & -\tilde\lambda_2 v_u &
-\tilde\lambda_3 v_u \\ 0 & 0 & 0 \\
\end{array}
\right].
\label{neutrino}
\eea
The block form displayed in (\ref{seesaw}) clearly demonstrates the
``see-saw structure'' of the mass matrix (\ref{nmatrix}).  We note
that the block matrix $m$ characterizes the lepton number violation in
the model.  Furthermore, the NMSSM with lepton number violation is
invariant under the $SU(4)$ group acting on the set of superfields
$(H_d, L_i)$, in the sense that the action of $SU(4)$ transformations
on $(H_d, L_i)$ leaves the superpotential form invariant up to
corresponding transformations of Yukawa couplings. We can use this
freedom to choose a basis which is characterised by vanishing
sneutrinos vacuum expectation values, $v_a =0$. We shall choose such a
basis in the following whenever it is convenient.

\begin{center} \begin{figure} \epsfxsize =6. in
\epsffile{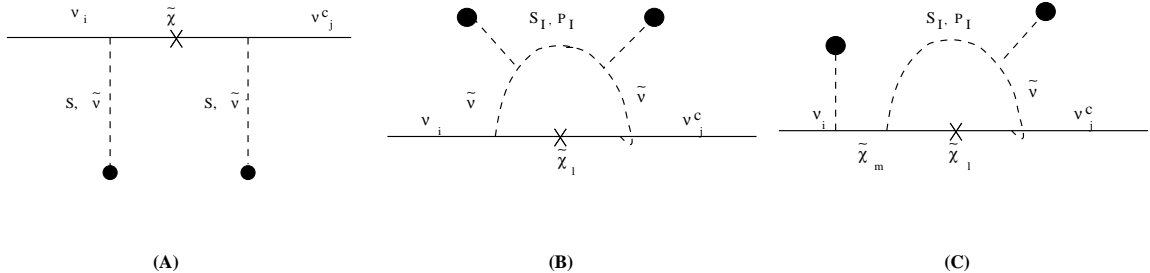}
\caption{The two point amplitudes for the process $\nu _i \to \nu ^c
_j $ associated with contributions to the neutrino Majorana mass
matrix at tree level (graph $(A)$)  due to exchange of neutralino
$\tchi _l $, and at one-loop level (graphs $(B) $ and $(C)$) due to
intermediate propagation of neutralino $\tchi _l $, sneutrino $\tilde
\nu _i $ and Higgs boson sector mass basis modes $ S_I, P_I$.  The
amplitude in $(A)$ is initiated by tadpoles (VEVs) of $ H_u, S, \tilde
\nu $,  and the amplitudes in $(B) $ and $(C)$ by double and single
tadpoles of $ H_u, H_d , S, \tilde \nu $.  The cross on the neutralino
propagator indicates a mass insertion term.}
\label{nmass} \end{figure} \end{center}

The masses of the neutralinos and neutrinos can be obtained by the
diagonalization of the mass matrix (\ref{nmatrix}) \bea {\cal N}^*
{\bf M_N} {\cal N}^{-1} & = & diag(m_{\chi^0_i}, m_{\nu_j}) \eea where
$m_{\chi^0_i}, (i = 1,....5)$ are the neutralino masses, and $
m_{\nu_j}, (j = 1, 2, 3)$ are the neutrino masses, respectively.  The
matrix (\ref{nmatrix}) cannot, in general, be diagonalized
analytically.  However, we are interested in the case where the tree
level neutrino masses as determined from the mass matrix
(\ref{nmatrix}) are small. In this case we can find approximate
analytical expression for the neutrino masses which are valid in the
limit of small lepton number violating couplings. To do so, we define
the matrix~\cite{schechter82} \bea \xi & = & m \cdot {\cal
M}_{\chi^0}^{-1}.
\label{defximatrix}
\eea If all the elements of this matrix are small, i. e.  \be \xi_{ij}
\ll 1,
\label{smallmatrix}
\ee then we can use it as an expansion parameter for finding an
approximate solution for the mixing matrix ${\cal N}$. Calculating the
matrix elements of $\xi_{ij}$ we find \bea \xi_{i1} & = & \frac{g_1
M_2 (\lambda x)}{\sqrt 2 \det ({\cal M}_{\chi^0})} [(2 \kappa x) x - 2
Y v_u] \Lambda_i, \nonumber \\ \xi_{i2} & = & - \frac{g_2 M_1 (\lambda
x)}{\sqrt 2 \det ({\cal M}_{\chi^0})} [(2 \kappa x) x - 2 Y v_u]
\Lambda_i, \nonumber \\ \xi_{i3} & = & - \frac{(g_2^2 M_1 + g_1^2
M_2)}{2 \det ({\cal M}_{\chi^0})} [(2\kappa x)v_d x + (\lambda v_u^2 -
Y v_d)v_u] \Lambda_i, \nonumber \\ \xi_{i4} & = & \frac{(g_2^2 M_1 +
g_1^2 M_2)}{2 \det ({\cal M}_{\chi^0})} (\lambda v_u^2 + Y v_d)x
\Lambda_i, \nonumber \\ \xi_{i5} & = & -\frac{\tilde \lambda_i
x}{\lambda x} [1 + \frac{(g_2^2 M_1 + g_1^2 M_2)}{2 \det ({\cal
M}_{\chi^0})} (\lambda v_u^2 - Y v_d)^2]\nonumber\\ && + \frac{(g_2^2
M_1 + g_1^2 M_2)}{2 \det ({\cal M}_{\chi^0})} [(2 \kappa x)v_u x
\Lambda_i + (\tilde\lambda_i v_u^2 -Y v_i)(\lambda v_u^2 -Y v_d)],
\label{defxielements}
\eea where we have used the notation \bea \Lambda_i & = & \lambda v_i
- \tilde \lambda_i v_d.
\label{deflambda}
\eea We note that $\xi_{i5}$ is not proportional to $\Lambda_i.$ From
Eqs.~(\ref{defxielements}) and (\ref {deflambda}), we see that $\xi =
0 $ in the MSSM limit where $ \tilde \lambda_i = 0, v_i = 0$.  The
matrix ${\cal N}^*$ which diagonalizes the neutralino-neutrino mass
matrix ${\bf M_N}$ can now be written as
\be {\cal N}^* \hskip -2pt =\hskip -2pt \left(\hskip -1pt
\begin{array}{cc}
N^* & 0\\ 0& V_\nu^T \end{array}
\hskip -1pt \right) \left(
\hskip -1pt
\begin{array}{cc}
1 -{1 \over 2} \xi^{\dagger} \xi& \xi^{\dagger} \\ -\xi & 1 -{1 \over
  2} \xi \xi^\dagger
\end{array}
\hskip -1pt \right),
\label{calnmatrix}
\ee
where we have retained only the leading order terms in $\xi$.  The
second matrix in (\ref{calnmatrix}) block diagonalizes the
neutralino-neutrino mass matrix ${\bf M}_N$ to the form diag~$({\cal
M}_{\chi^0}, m_{eff})$, with \bea m_{eff} & = & - m \cdot {\cal
M}_{\chi^0}^{-1} \cdot m^T \cr \vb{18} & = & \frac{(M_1 g^2 \!+\! M_2
{g'}^2) (\kappa x^2 - Y v_u) x} {det({\cal M}_{\chi^0})} \left(\hskip
-2mm \begin{array}{ccc} \Lambda_e^2
\hskip -1pt&\hskip -1pt \Lambda_e \Lambda_\mu
\hskip -1pt&\hskip -1pt \Lambda_e \Lambda_\tau \\ \Lambda_e
\Lambda_\mu
\hskip -1pt&\hskip -1pt \Lambda_\mu^2
\hskip -1pt&\hskip -1pt \Lambda_\mu \Lambda_\tau \\ \Lambda_e
\Lambda_\tau
\hskip -1pt&\hskip -1pt \Lambda_\mu \Lambda_\tau
\hskip -1pt&\hskip -1pt \Lambda_\tau^2
\end{array}\hskip -3mm \right),
\label{defmeff}
\eea where the tree level contribution to the light neutrino mass
matrix admits the Feynman diagram representation of
Figure~\ref{nmass}(A). The eigenvalues of $m_{eff}$ give the tree
level neutrino masses.  These eigenvalues are \bea m_{\nu _3} & = &
\frac{(M_1 g^2 \!+\!  M_2 {g'}^2) (\kappa x^2 - Y v_u) x} {det({\cal
M}_{\chi^0})}\sum_i \Lambda_i^2
\label{numass3}, \\
m_{\nu _1} & = & m_{\nu _2} = 0,
\label{numass12}
\eea
\vskip .5 cm
where we have used $ m_{\nu _3} \ge m_{\nu _2} \ge m_{\nu _1}.$ Thus,
at the tree level only one neutrino is massive. Its mass is
proportional to the lepton number~(and $R$-parity) violating parameter
$\sum_i \Lambda_i^2$. A single lepton number violating coupling
$\tilde \lambda_i$ can lead to a non-zero neutrino mass.  Furthermore,
we note that the sub-matrices $N$ and $V_{\nu}$ diagonalize ${\cal
M}_{\chi^0}$ and $m_{eff}$: \bea N^{*}{\cal M}_{\chi^0} N^{\dagger} =
{\rm diag}(m_{\chi^0_i}), \\ V_{\nu}^T m_{eff} V_{\nu} = {\rm
diag}(0,0,m_{\nu _3}).  \eea Since only one of the neutrinos obtains
mass, we can rotate away one of the three angles in the matrix
$V_{\nu}$. We can then write $V_{\nu}$ as a product of two
matrices~\cite{schechter80} \be V_{\nu}= \left(\begin{array}{ccc} 1 &
0 & 0 \\ 0 & \cos\theta_{23} & -\sin\theta_{23} \\ 0 & \sin\theta_{23}
& \cos\theta_{23}
\end{array}\right) 
\left(\begin{array}{ccc} \cos\theta_{13} & 0 & -\sin\theta_{13} \\ 0 &
  1 & 0 \\ \sin\theta_{13} & 0 & \cos\theta_{13}
\end{array}\right) ,
\ee where the mixing angles can be written in terms of $\Lambda_i$ as
\bea \tan\theta_{13} & = & \frac{\Lambda_e}
{(\Lambda_{\mu}^2+\Lambda_{\tau}^2)^{\frac{1}{2}}} ,
\label{theta13}\\
\tan\theta_{23} & = & - \frac{\Lambda_{\mu}}{\Lambda_{\tau}}.
\label{theta23}
\eea
Finally, using the expression \bea det({\cal M}_{\chi^0}) & = & (2
\kappa x)(\lambda x) [(g_1^2M_2 + g_2^2 M_1) v_u v_d - M_1 M_2
(\lambda x )^2] + 2 M_1 M_2 (\lambda v_u) (\lambda x) Y \nonumber \\ &
+ & \frac {1}{2} (g_1^2 M_1 + g_2^2 M_2) (\lambda v_u^2 - v_d Y)^2,
\label{determinant}
\eea for the determinant of ${\cal M}_{\chi^0}$ in (\ref{numass3}), we
have for the tree level mass of the neutrino \bea m_{\nu _3} & \sim &
\frac {\cos^2\beta}{\tilde m} \sum_i (\tilde \lambda_i x)^2,
\label{m3value}
\eea where we have assumed that all the relevant masses~(and the
relevant vacuum expectation values) are at the electroweak~(or
supersymmetry breaking scale) scale denoted by $\tilde m$.  For
simplicity we have chosen the basis in which the sneutrino vacuum
expectation values $v_i = 0$ to write the result (\ref{m3value}).  We,
thus, see that apart from the $R$-parity violating parameter $\sum_i
(\tilde \lambda_i x)^2$, the tree level neutrino mass is proportional
to $\cos^2\beta$. For large values of $\tan\beta$, this could lead to
a suppression of $m_{\nu _3}$, which could be important.

It is now important to calculate the admixture of the singlet
component~(arising from the fermionic component of the Higgs singlet
superfield $S$) in the three light neutrino states. From
(\ref{calnmatrix}) we can write the matrix ${\cal N}^*$ which
diagonalizes the neutralino-neutrino mass matrix as \be {\cal N}^*
\hskip -2pt =\hskip -2pt \left(\hskip -1pt
\begin{array}{cc}
N^*(1 -{1 \over 2} \xi^{\dagger} \xi)& N^* \xi^{\dagger}\\ -
V_{\nu}^T\xi & V_\nu^T(1 - {1 \over 2} \xi \xi^{\dagger}) \end{array}
\hskip -1pt \right).
\label{calnmatrix1}
\ee The eigenvectors of the neutralino-neutrino mass matrix are then
given by \be F_i^0 = {\cal N}_{ij} \psi_j,
\label{eigenvectors}
\ee where as indicated above we use the basis $\psi_j = (-i \lambda',
-i \lambda^3, \tilde H_u, \tilde S, \tilde H_d, \nu _e, \nu _{\mu},
\nu _{\tau}).$ The singlet component in the three neutrino states is
then given by $|{\cal N}_{64}|^2, |{\cal N}_{74}|^2$ and $|{\cal
  N}_{84}|^2,$ respectively.  For calculating these components we
require the sub-matrix $ V_{\nu}^T \xi $ of the matrix
(\ref{calnmatrix1}). It is straightforward to calculate this
sub-matrix, and the result is~($\vec\Lambda = (\Lambda_e,
\Lambda_{\mu}, \Lambda_{\tau}$)) \be V_{\nu}^T \xi \hskip -2pt =\hskip
-2pt \left(\hskip -1pt
\begin{array}{ccccc}
0 & 0 & 0 & 0 & \tilde \epsilon_1\\ 0 & 0 & 0 & 0 & \tilde
\epsilon_2\\ a_1 |\vec\Lambda| & a_2 |\vec\Lambda| & a_3 |\vec\Lambda|
& a_4 |\vec\Lambda| & \tilde \epsilon_3 + a_5 |\vec\Lambda|
\hskip -1pt
\end{array}
\right),
\label{submatrix}
\ee where \bea \tilde \epsilon_1 & = & \frac{\tilde \epsilon_e
(\Lambda_{\mu}^2 + \Lambda_{\tau}^2) - \Lambda_e(\Lambda_{\mu} \tilde
\epsilon_{\mu} + \Lambda_{\tau} \tilde \epsilon_{\tau})}
{\sqrt{(\Lambda_{\mu}^2 + \Lambda_{\tau}^2)} \sqrt{(\Lambda_e^2 +
\Lambda_{\mu}^2 + \Lambda_{\tau}^2)}}, \\
\tilde \epsilon_2 & = & \frac{- \tilde \epsilon_{\mu} \Lambda_{\tau} +
\tilde \epsilon_{\tau} \Lambda_{\mu}} {\sqrt{(\Lambda_{\mu}^2 +
\Lambda_{\tau}^2)}},\\
\tilde \epsilon_3 & = & - \frac{\vec\Lambda\cdot\vec{\tilde\epsilon}}
{\sqrt{(\Lambda_e^2 + \Lambda_{\mu}^2 + \Lambda_{\tau}^2)}},\\
\tilde\epsilon_e & = & -\frac{\tilde \lambda_1 x}{\lambda x} [1 +
\frac{(g_2^2 M_1 + g_1^2 M_2)}{2 \det ({\cal M}_{\chi^0})} (\lambda
v_u^2 - Y v_d)^2]\nonumber\\ && + \frac{(g_2^2 M_1 + g_1^2 M_2)}{2
\det ({\cal M}_{\chi^0})} [(\tilde\lambda_i v_u^2 -Y v_1)(\lambda
v_u^2 -Y v_d)],\\
\tilde\epsilon_{\mu} & = & -\frac{\tilde \lambda_2 x}{\lambda x} [1 +
\frac{(g_2^2 M_1 + g_1^2 M_2)}{2 \det ({\cal M}_{\chi^0})} (\lambda
v_u^2 - Y v_d)^2]\nonumber\\ && + \frac{(g_2^2 M_1 + g_1^2 M_2)}{2
\det ({\cal M}_{\chi^0})} [(\tilde\lambda_2 v_u^2 -Y v_2)(\lambda
v_u^2 -Y v_d)],\\
\tilde\epsilon_{\tau} & = & -\frac{\tilde \lambda_3 x}{\lambda x} [1 +
\frac{(g_2^2 M_1 + g_1^2 M_2)}{2 \det ({\cal M}_{\chi^0})} (\lambda
v_u^2 - Y v_d)^2]\nonumber\\ && + \frac{(g_2^2 M_1 + g_1^2 M_2)}{2
\det ({\cal M}_{\chi^0})} [(\tilde\lambda_3 v_u^2 -Y v_3)(\lambda
v_u^2 -Y v_d)],\\
\eea and \bea a_1 & = & - \frac{g_1 M_2 (\lambda x)}{\sqrt 2 \det
({\cal M}_{\chi^0})} [(2 \kappa x) x - 2 Y v_u], \\ a_2 & = &
\frac{g_2 M_1 (\lambda x)}{\sqrt 2 \det ({\cal M}_{\chi^0})} [(2
\kappa x) x - 2 Y v_u], \\ a_3 & = & \frac{(g_2^2 M_1 + g_1^2 M_2)}{2
\det ({\cal M}_{\chi^0})} [(2\kappa x)v_d x + (\lambda v_u^2 - Y
v_d)v_u], \\ a_4 & = & \frac{(g_2^2 M_1 + g_1^2 M_2)}{2 \det ({\cal
M}_{\chi^0})} (\lambda v_u^2 + Y v_d)x, \\ a_5 & = & - \frac{(g_2^2
M_1 + g_1^2 M_2)}{2 \det ({\cal M}_{\chi^0})} [(2 \kappa x)v_u
x]. \eea From (\ref{calnmatrix1}) and (\ref{submatrix}) we obtain the
important result \bea |{\cal N}_{64}|^2 & = & |{\cal N}_{74}|^2 = 0,
\\ |{\cal N}_{84}|^2 & = & a_4^2 |\vec\Lambda|^2. \label{mixture} \eea
Thus, at the tree level two light neutrinos do not have a singlet
component, whereas the heaviest neutrino has a singlet component with
a strength proportional to the square of the lepton number violating
parameter $|\vec{\Lambda}|.$

\subsection{\bf One-loop  supersymmetry breaking  contributions}
\label{secloop}

As shown above, at the tree level only one of the neutrinos obtains a
mass through the lepton number violating Yukawa coupling $\tilde
\lambda_i,$ so that the tree level neutrino mass matrix can be written
as \be m_{\nu}^0 \equiv V_{\nu}^T m_{eff} V_{\nu} = {\rm
diag}(0,0,m_{\nu _3}). \label{numatrix} \ee However, the neutrino mass
matrix can receive contributions from loop effects.
The supersymmetry breaking parameters are expected to play a crucial
r\^ole through the one-loop corrections involving gauge interactions
with exchange of sneutrinos~\cite{grosshaber99,davidson00}.  At
one-loop level the needed suppression of neutrino masses can arise
from cancellations between contributions involving the Higgs sector,
and from possible mass degeneracies among the sneutrinos. In the
context of MSSM, this has been discussed in~\cite{grossm04}.

At one-loop level, finite Majorana neutrino masses can be generated
through two classes of mechanisms involving either the gauge or
superpotential interactions in combination with the soft supersymmetry
breaking interactions.  These mechanisms have been discussed in detail
for the MSSM~\cite{chemtob}.  The loop amplitudes in the former class
propagate matter particles and contribute at orders $ \tilde \l _i
\tilde \l_j,\ \tilde \l _i \l _{ijk} $ and $\tilde \l _i \l '_{ijk} $,
and those in the latter class propagate sleptons and gauginos and
contribute at orders $ A_ {\tilde \l _i} \tilde \l _i A_ {\tilde \l
_j} \tilde \l _j ,\ A_ {\tilde \l _i} \tilde \l _i \tilde \l _j $.
These are associated with the mixing of sneutrinos and Higgs bosons, 
and are also responsible for the mass splittings between
sneutrinos and antisneutrinos~\cite{grosshaber99}.  The possibility
that the combined tree and one-loop contributions in the MSSM could
account for the observed flavor hierarchies in the masses and mixing
angles of light neutrinos has been studied in several recent
works~\cite{davidson00,abada02,borzum02,chun02,grossm04, valle03}.  The
supersymmetry breaking interactions, the cancellations between
contributions involving the Higgs sector modes, and mass splittings
among sneutrinos of different flavors, are expected to play a crucial
r\^ole.

The present section is aimed at studying the one-loop contributions
to the neutrino mass matrix, and the extent to which these constitute a
sensitive probe of the Higgs boson sector of the NMSSM.  The finite
VEVs for the components of the scalar fields $ H_d, H_u, S, 
\tilde \nu _i $ can result in one-loop contributions for the two-point 
amplitude represented by the Majorana mass term 
$ L_{EFF}= - \ud (m_{\nu })_{ij} \bar \nu ^c _j\nu _i + H.\ c. $. 
These contributions are displayed by the Feynman diagrams in Fig.\ref{nmass}(B) and (C), with double and
single mass sneutrino-scalar mass mixing insertion terms.  These are
the analogs of MSSM~\cite{grosshaber99} for the case of NMSSM.  It is
important to carefully treat these see-saw like contributions by
expressing the intermediate scalar and pseudoscalar propagators in the
mass eigenbasis.  We shall first obtain the scalar potential for the
sneutrinos and Higgs bosons, minimize it with respect to the
corresponding VEVs, $ v_d, v_u, x, v _i $, extract the squared mass
matrix whose off-diagonal blocks represent the sneutrino-Higgs mass
mixing terms, and finally evaluate the one-loop contributions to the
neutrino Majorana mass terms, $ (m_{\nu })_{ij} $.

\subsubsection{\bf Coupling of Higgs boson  and sneutrino sectors}

Using the standard procedure, we can write down the scalar potential
of the NMSSM involving the relevant components of the complex scalar
fields in terms of the $F$-terms, arising from the superpotential, the
$D$-terms, arising from the gauge interactions, and the soft
supersymmetry breaking terms as follows: \bea W_{\nu} & = & \l H _d
H_u S + \tilde \l _i L_i H_u S - {\kappa \over 3} S^3, \nonumber \\ V
& = & V_F + V_D + V_{soft},\nonumber \\ V_F & = &\sum_i \vert \frac{
\dh W_{\nu}} {\dh \phi_i} \vert ^2, ~~~~~~~ V_D = {g_1^2 + g_2^{2}
\over 8 } ( \vert v_u \vert ^2 - \vert v _A\vert ^2 ) ^2, \nonumber \\
V_{soft} & = & \sum _{AB} M^2 _{\tilde \nu _ B \tilde \nu _ A} \tilde
\nu ^\star _B \tilde \nu _A + m_{H_u} ^2 \vert v_u \vert ^2 + m_S ^2
\vert x \vert ^2\nonumber\\ && - [ A _{\tilde \l _A } \tilde \l _A v
_A v_u + {A _{\kappa }\kappa \over 3 } x^3 + H.\ c. ].
\label{potential}
\eea In (\ref{potential}), $\phi_i$ stand for all the relevant scalar
fields, and we have used the convenient four-vector notations $ \tilde
\l_A = (\l , \tilde \l _i), \ v _A = (v_d, v_i )$, with the summation
convention over repeated indices undestood. The VEVs of different
fields are temporarily extended to complex numbers, $ v_d = <H_d> =
v_{d1} + i v _{d2}, \ v_u = <H_u> = v_{u1} + i v_{u2}, \ x =<S> = x_1
+i x_2,\ v _i =<\tilde \nu _i > = v_{i1} + i v _{i2} $, corresponding
to the decomposition of Higgs boson and sneutrino fields into real
scalar $CP$-even and imaginary pseudoscalar $CP$-odd
components~\cite{ellis89} \bea && H_d=
{H_{d R} + i H_{d I} \over \sqrt 2 }, \ H_u=
{H_{u R} + i H_{u I} \over \sqrt 2 },\ S=
{S_{R} + i S_{I} \over \sqrt 2 } , \ \tilde \nu _{i} = {\nu _{iR} +i
\nu _{iI} \over \sqrt 2 }.\eea This basis for the scalar fields $
(H_{d R} , H_{u R} , S_{R} ) ,\ (H_{d I} , H_{u I} , S_{I} ) $ is
related to the mass eigenstate basis $ (S_I), \ (P_I) , \ [I=1,2,3] $
by the linear transformations \bea \pmatrix{H_{d R} \cr H_{u R} \cr
S_{R} } - \sqrt 2 \Re \pmatrix{v_d \cr v_u \cr x} &=& U _s^T
\pmatrix{S_1 \cr S_2 \cr S_3} , \ \pmatrix{H_{d I} \cr H_{u I} \cr
S_{I} } - \sqrt 2 \Im \pmatrix{v_d \cr v_u \cr x} = U _p^T
\pmatrix{P_1 \cr P_2 \cr P_3} , \eea where $ U_s ,\ U_p $ denote the
unitary matrices which diagonalize the mass squared matrices in the
interaction basis, $ (M^2 )_{H_{dR}, H_{uR}, S_R } \equiv M^2 _{s, ij}$
and $ (M^2 )_{H_{dI}, H_{uI}, S_I } \equiv M^2 _{p, ij} $, using the
definition $ U_{s,p} ^T M^2 _{s,p} U_{s,p} = (M^2 _{s,p})_{diag}.$ We
then minimize the scalar potential with respect to the VEVs of various
fields, and eliminate the soft supersymmetry breaking mass parameters
$ m^2 _{H_u}, \ m^2 _{S}, \ m^2 _{H_d} \ M^2 _{\tilde \nu _i \tilde
\nu _j} v_j ^\star $ through the equations \bea {\dh V \over \dh v_u }
& = & 0,\ {\dh V \over \dh x }= 0,\ {\dh V \over \dh v _A }\equiv \ud
( {\dh V \over \dh v _{A R} } -i {\dh V \over \dh v _{A I} } ) =
0. \eea The mass squared matrices for the $CP$-even and $CP$-odd sector
fields $(H_d, H_u, S,\tilde \nu _i)_{R,I}$ are then evaluated by
applying the definitions \bea M^ {2} _{s,ij} & = & {d^2 V \over \dh
\phi _{i R} \dh \phi _{j R} } ,\ M^ {2} _{p,ij} = {d^2 V \over \dh
\phi _{iI} \dh \phi _{j I} }.  \eea Finally, we restrict our
considerations to the physical vacuum solutions with vanishing
imaginary parts of the field VEVs, $ \Im (v_u )=0, \ \Im (v_d )=0, \
\Im (x )=0, \ \Im ( v_i )=0$.  For convenience, and without loss of
generality, we shall also specialize to the choice of $L_A$ field
basis characterized by vanishing complex sneutrino VEVs, $ v_i =0$.
While feasible, the basis independent analysis in the supersymmetry
breaking case~\cite{davidson00,haberbasis} is significantly
complicated by the need to account for several independent algebraic
invariants.  As a function of the dimensionless parameters $ (\l , \
\tilde \l _i , \ \kappa )$, of the dimensional supersymmetry breaking
parameters $ (A _\l , \ A_{\tilde \l _i} ,\ A_ \kappa ) $, which
include the gravitino mass parameter $ m_{3/2}$, and of the scalar
field VEVs $(v_{d},\ v_{u}, \ x)$, the scalar and pseudoscalar mass
squared matrices $M^2 _{s,ij} ,\ M^2 _{p, ij},\ [i, j = d,u,S,\tilde
\nu _k]$ are given by the symmetric matrices \bea M^2 _{s, dd} & = &
{1\over v_d} [ {g_2^2 + g_1^2 \over 2} v_d ^3 + v_u x (A _\l \l +
\kappa \l x) ] ,\ M^2 _{s, du} = -{g_2^2 + g_1^2 \over 2} v_d v_u + 2
\l ^2 v_d v_u - A _\l \l x - \kappa \l ^2 x ^2 , \cr && M^2 _{s, dS} =
-(A _\l \l v_u) + 2 \l (\l v_d - \kappa v_u) x,\ M^2 _{s, d\tilde \nu
_i} = { v_u x\over v_d} (A_{\tilde \l _i} \tilde \l _i + \kappa \tilde
\l _i x) , \cr && M^2 _{s, uu} = {1\over v_u} [{g_2^2 + g_1^2 \over 2}
v_u ^ 3 + v_d x (A _\l \l + \kappa \l x) ], \ M^2 _{s, uS} = - A _\l
\l v_d + 2 [ -\kappa \l v_d + (\l ^2 + \tilde \l _i ^2 ) v_u] x,\cr &&
M^2 _{s, u\tilde \nu _i} = 2 \l \tilde \l _i v_d v_u - x (A_{\tilde \l
_i} \tilde \l _i + \kappa \tilde \l _i x) , \cr && M^2 _{s, SS} ={A
_\l \l v_d v_u \over x } + x (-A _{\kappa } \kappa + 4 \kappa ^2 x), \
M^2 _{s, S\tilde \nu _i} = - A_{\l _i} \tilde \l _i v_u + 2 \tilde \l
_i (\l v_d - \kappa v_u) x , \cr && M^2 _{s, \tilde \nu _i \tilde \nu
_j } = M^2 _{s, \tilde \nu \tilde \nu } + {g_2^2 + g_1^2 \over 4}
(v_d^2 - v_u ^2 ) + \tilde \l _i ^2 (v_u ^2 + x ^2),
\label{eqmass1} \eea 
for the $CP$-even scalars,  and the symmetric matrices \bea && M^2 _{p,
dd} = { v_u x \over v_d} (A _\l \l + \kappa \l x),\ M^2 _{p, du} = x
(A _\l \l + \kappa \l x),\cr && M^2 _{p, dS} = v_u (A _\l \l - 2
\kappa \l x), \ M^2 _{p, d\tilde \nu _i} = { v_u x \over v_d}
(A_{\tilde \l _i} \tilde \l _i + \kappa \tilde \l _i x), \cr && M^2
_{p, uu} = { v_d x\over v_u} (A _\l \l + \kappa \l x),\ \ M^2 _{p, uS}
= v_d (A _\l \l - 2 \kappa \l x),\ M^2 _{p, u\tilde \nu _i} = x
(A_{\tilde \l _i} \tilde \l _i + \kappa \tilde \l _i x),\cr && M^2
_{p, SS} = 4 \kappa \l v_d v_u + { A _\l \l v_d v_u \over x } + 3 A
_{\kappa } \kappa x,\ M^2 _{p, S\tilde \nu _i} = v_u (A_{\tilde \l _i}
\tilde \l _i - 2 \kappa \tilde \l _i x),\cr && M^2 _{p, \tilde \nu _i
\tilde \nu _i} = M^2 _{\tilde \nu _i \tilde \nu _j } + {g_2^2 + g_1^2
\over 4 } (v_d ^2 -v_u^2 ) + \tilde \l _i ^2 (v_u ^2 + x^2 ) ,
\label{eqmass2} \eea for the $CP$-odd scalars.  The orthogonal linear
combinations of $CP$-odd scalar fields, $G^0 (x) = \cos \b H_{d I} (x) -
\sin \b H_{u I} (x) ,\ A (x) = \sin \b H_{d I} (x) + \cos \b H_{u I}
(x) $,
identified with the decoupled Goldstone field which is absorbed as the
longitudinal polarization mode of $Z^0$ and with the axionic symmetry
pseudo-Goldstone boson mode $A$, respectively.  The mass squared
matrix in the field basis $ [A, S_I, \tilde \nu _i]$ is obtained by
first applying the similarity transformation $(G ^0, A, S_I) ^T =
\calr ^T (H_{dI}, H_{uI}, S_I ) ^T,\ [\calr = diag (\calr _\b , 1) ] $
with $\calr _\b $ denoting the $ SO(2)$ rotation of angle $\b $, and
next by removing the decoupled Goldstone mode $G ^0$. The mass squared
matrix in the transformed basis, $ (M^2 )_{G, A , S _I} = \calr ^T
(M^2 )_{H_{dI}, H_{uI}, S_I } \calr ,\ [\calr = diag (\calr _\b , 1_4)
] $ can then be written as \bea && M^2 _{p, AA}= \sin ^2 \b M^2
_{p,dd} + \cos ^2 \b M^2 _{p,uu} +2 \sin \b \cos \b M^2 _{p,ud} =
{1\over \cos \b \sin \b } x (A_\l \l + \kappa \l x) , \cr && M^2 _{p,
AS}= \sin \b M^2 _{p,dS} + \cos \b M^2 _{p,uS} = v (A_\l \l -2 \kappa
\l x) , \cr && M^2 _{p, A \tilde \nu _i }= {x\over \cos \b } (A
_{\tilde \l _i} \tilde \l _i + \kappa \tilde \l _i x) . \eea These
results, with finite $\tilde \l _i x $, are a generalization of the
results of NMSSM~\cite{ellis89} to the case when there is lepton
number violation induced by trilinear couplings. These results reduce
to the corresponding results of MSSM with lepton number
violation~\cite{grosshaber99} in the limit $x \to \infty $ with fixed
$ \l x = -\mu ,\ \tilde \l _i x = -\mu _i, $\ and $\kappa x ^3 $.

Since the off-diagonal entries in the sneutrino-Higgs mass matrix can
be safely assumed to be small in comparison to the diagonal entries,
one may evaluate the contributions to the sneutrino mass splittings by
making use of second order matrix perturbation theory.  The same
approximation is also used in evaluating the modified sneutrino
propagators in the mass insertion approximation.  Specifically, the
Higgs boson sector propagator, modified by tbe two mass mixing terms
$\tilde \nu _i \times (S_I \oplus P_I) \times \tilde \nu _j $ in the
Feynman diagram of Fig.~\ref{nmass}(B), can be represented by the
weighted propagator \bea P_{\tilde \nu _i \tilde \nu _j} (q) & \equiv
& \sum _{J=1,2,3} { s ^J_{ij} \over q ^2 - M^2 _{S_J} } - \sum
_{J=1,2} {p _{ij} ^J \over q ^2 - M^2 _{P_J} }, \eea where \bea s_{ij}
^J & = & \bigg ( \sum _{k = d,u,S } M^2 _{s, \tilde \nu _{i} k } (U_s
^T)_{kJ} \bigg ) \bigg ( \sum _{k = d,u,S} M^2 _{s, \tilde \nu _{j} k
} (U_s ^T)_{kJ} \bigg ),\\ p _{ij} ^J & = & \bigg ( \sum _{k = d,u,S}
M^2 _{p, \tilde \nu _{i} k } (U_p ^T)_{kJ} \bigg ) \bigg ( \sum _{k =
d,u,S} M^2 _{p, \tilde \nu _{i} k } (U_p ^T)_{kJ} \bigg ).  \eea
Evaluating the transition amplitude for the double mass insertion
one-loop Feynman graph of Fig.~\ref{nmass}(B) with the above formula
for the weighted Higgs boson propagator, one obtains the contribution
to the light neutrino mass matrix \bea (m_\nu )_{ij} ^B & = &
{g_2^2\over 4 } \sum _l M_{\tchi _l } (N_{l2} - \tan \t _W N_{l1} ) ^2
[ \sum _{J=1,2,3} s _{ij} ^J I_{4} ( m _{\tilde \nu _i}, m _{\tilde
\nu _j} M_{\tchi _l }, M _{S_J} ) \nonumber \\ && - \sum _{J=1,2} p
_{ij} ^J I_{4} ( m _{\tilde \nu _i}, m _{\tilde \nu _j}, M_{\tchi _l
}, M _{P_J} ) ], \eea with \bea [I_{4} ( m _{\tilde \nu _i}, m
_{\tilde \nu _j}, M_{\tchi _l }, M _{X_J} ) & \equiv & { 1 \over (4\pi
)^2 } C ( m _{\tilde \nu _i}, m _{\tilde \nu _j}, M_{\tchi _l }, M
_{X_J} ) \nonumber\\ & = & \int {d^4 q\over i (2\pi )^4 } {1\over (q^2
- m^2 _{\tilde \nu _i } )(q^2 - m^2 _{\tilde \nu _j } ) (q^2 -
M_{\tchi _l } ^2 ) (q ^2 - M^2 _{X_J})}], \nonumber\\ \eea where $\tan
\t _W = g_1/g_2$, and we have used the matrix $N$ to denote the
unitary transformation linking the interaction and mass eigenstates of
massive neutralinos, $ (\tchi _m) _{mass}= (N ^\dagger ) _{ml}\tchi
_l$.  The momentum integral $I_4$ admits the analytic
representation~\cite{passvel79} \bea I_4 (m_1, m_2,m_3,m_4) & = & {1
\over m_3 ^2 - m_4 ^2} [I_3 (m_1, m_2,m_3) - I_3 (m_1, m_2,m_4)] ,\cr
I_3 (m_1, m_2,m_3) & = & {1 \over m_2^2- m_3^2} [I_2 (m_1, m_2) - I_3
(m_1, m_3)] , \cr I_2 (m_1, m_2) & = & {1\over (4\pi )^2} { m_1 ^2
\over m_2 ^2 - m_1^2 } \log {m_1^2 \over m_2^2}. \eea The single mass
insertion one-loop amplitude, displayed in Feynman graph of
Fig.~\ref{nmass}(C), yields the following contribution to the light
neutrino mass matrix \bea ( m_\nu )_{ij} ^C & = & {g_2^2 \over 4 }
\sum _{l,m} { M_{\tchi _m } \over M_{\tchi _l} } \tilde \l _i x \bigg
[ \bigg ( N_{l4} (N_{m2} - \tan \t _W N_{m1} ) + N_{m4} (N_{l2} - \tan
\t _W N_{l1} ) \bigg ) \nonumber \\ && \times [ \sum _{J=1}^3 (U_s ^T)
_{dJ} Q ^s_{\tilde \nu _j J} - \sum _{J=1}^2 (U_p ^T) _{dJ} Q ^p_{
\tilde \nu _j J} ] \nonumber \\ && - N_{l3} (N_{m2} - \tan \t _W
N_{m1} ) [ \sum _{J=1}^3 (U_s ^T) _{uJ} Q ^s_{ \tilde \nu _j J} - \sum
_{J=1}^2 (U_p ^T) _{uJ} Q^p_{ \tilde \nu _j J} ] \bigg ] + (i
\leftrightarrow j ), \nonumber \\ \eea where various quantities in the
above equation are defined as \bea Q^s_{ \tilde \nu _j J} & = &
\sum_{k= d, u, S } M^2 _{s.  \tilde \nu _j k } (U_s ^T) _{kJ} I_3 (
m_{\tilde \nu _j } , M_{\tchi _m}, M _{S_J} ),\\ Q^p_{ \tilde \nu _j
J} & = & \sum_{k= d, u, S } M^2 _{p. \tilde \nu _j k } (U_s ^T) _{kJ}
I_3 ( m_{\tilde \nu _i } , M_{\tchi _m} , M^2 _{P_J} ),\\ I_{3} ( m
_{\tilde \nu _i}, M_{\tchi _m}, M _{X_J} ) & \equiv & { 1 \over (4\pi
)^2 } C _3 ( m _{\tilde \nu _i}, M_{\tchi _m}, M _{X_J} ) \\ & = &
\int { d^4 q \over i (2\pi )^4 } {1\over (q^2 - m^2 _{\tilde \nu _i }
)(q^2 - M^2 _{\tchi _m } ) (q ^2 - M^2 _{X_J}) } .\eea


An examination of the off-diagonal matrix elements of the
sneutrino-Higgs squared mass matrix given by equations (\ref{eqmass1})
and (\ref{eqmass2}) shows that the above one-loop contributions to
the neutrino mass matrix consist of sums of two separate matrices
involving the three-vectors, $ \tilde \l _i $ and $A_{\tilde \l
_i}\tilde \l _i $ in the space of fields $L_i $.  Combining these with
the tree contribution discussed in subsection~\ref{sectree}, one can
now write the following representation of the effective light neutrino
mass matrix as a sum of three contributions \bea (m_\nu )_{ij} & = &
(X_A ^t + X_B ^l +X_C ^l ) \tilde \l _i \tilde \l _j + Y_B ^l
A_{\tilde \l _i} \tilde \l _i A_{\tilde \l _j} \tilde \l _j\nonumber\\
&& + (Z_B ^l +Z_C ^l ) (\tilde \l _i A_{\tilde \l _j} \tilde \l _j +
A_{\tilde \l _i} \tilde \l _i \tilde \l _j ), \eea where the lower
suffix labels $A$, and $ B, C$ in the coefficients $ X, Y, Z$ refer to
the tree and one-loop contributions coming from the Feynman diagrams
$(A)$, and $( B ), \ (C)$ in Fig.~\ref{nmass} and we have appended the
upper suffix labels $t , \ l$ to emphasize the distinction between
tree and one-loop contributions.  We note the absence of the
coefficient $Y_C ^l$, and the relation $ X_B ^l, X_C ^l << X_A ^t $
expected from the loop suppression factor, which allows us to ignore
the coefficients $X _B$ and $X_C ^l $.  The single mass insertion
contributions $Z_B ^l, \ Z_C ^l$, and the double mass insertion term
$Y_B ^l$ have the ability, either separately or in combination, to
produce a second non-vanishing mass eigenvalue, provided only that the
three-vector $A_{\tilde \l _i} \tilde \l _i $ is not aligned with $ \l
_i$.  We recall that the three-vector proportionality, $A_{\tilde \l
_i} \tilde \l _i \propto \l _i$, would hold if supersymmetry breaking
were flavor universal.  Moreover, as was first observed by Chun et
al.~\cite{chun02} in the context of MSSM, application of matrix
perturbation theory to the additively separable neutrino mass matrix
$(m_\nu )_{ij} = x \mu _i \mu _j + y b_i b_j + z (\mu _i b_j + \mu _j
b_i) $ indicates that the two finite eigenvalues present in the limit
$ y, z << x $ are given by $x \mu _i ^2 + 2 y b_i \mu _i + O(y^2,
z^2)$ and $ y b _i ^2 + O(y^2, z ^2) $.  Hence, assuming in the above
Majorana neutrino mass matrix $(m_\nu )_{ij}$ that the coefficients $
Y _B ^l , \ Z_{B,C} ^l $ are of subleading order relative to $ X_A ^t
$, one concludes that the second non-vanishing eigenvalue is of first
order in $ Y_B ^l $ but of second order in $ Z_B ^l,\ Z_C ^l $,
namely, $ m_{\nu _2} \simeq O(Y_B ^l) + O( Z_B ^{l2}, Z_C ^{l2}) $.
Thus, as far as the second mass eigenvalue is concerned, this implies
that the single mass insertion amplitude $(C)$ is subdominant, so that
we can restrict consideration to the double mass insertion
contribution $(B)$ only.

A rough estimate of various contributions can now be obtained by
isolating the stronger dependence on $\tan \b $, while assuming
that all the mass parameters take values of same order of magnitude as
the supersymmetry breaking mass scale $\tilde m_0 $. This yields the
approximate formulas for the coefficients representing the tree and
one-loop contributions, $ X_A ^t$ and $ Y_B ^l,\ Z_B^l,\ Z_C^l$, 
respectively: \bea
X ^t _A & \simeq & { x^2 \cos ^2 \b \over \tilde m _0 } ,\
Y_B ^l \simeq {x^2\over \tilde m _0 \cos ^2 \b } \e _L\e _H,
\nonumber\\ Z _B ^l & \simeq& {\kappa x^2 \over \tilde m_0 \cos ^2 \b
} \e _L \e _H , \ Z_C ^l \simeq {\kappa x^2 \over \tilde m_0 \cos \b }
\e _L \e '_H , \eea where we have included the suppression effect from
the loop in the factor $\e _L \simeq 1/(4\pi )^2 \sim 10^{-2} $ and
that from the Higgs sector in the factors $\e _H$ and $\e '_H$.  Note
that we have omitted the one-loop contributions to the component
$\tilde \l _i \tilde \l _j$, which are associated with  the suppressed
coefficients, $ X_B ^l \simeq Z_B ^l, \ X_C ^l \simeq Z_C ^l.$
The Higgs sector decoupling effect arises from the cancellation
between the contributions from $CP$-even and $CP$-odd scalars, and is most
effective in the case where the lightest scalar contribution is well
separated from the other modes, and the mass spectrum is ordered as, $
m_{S_1} \simeq m_Z << m_{S_2} , m_{S_3}, m_{P_1}, m_{P_2} .$ The
dominant contribution to the second finite neutrino mass eigenvalue
is, then, of order $ m_{\nu _2} \simeq <A{\tilde \l } ^2\tilde \l ^2 >
\cos ^2 \b \e _L \e _H /\tilde m_0 $.  Moreover, a third finite mass
eigenvalue may be generated from the one-loop amplitudes under study
by taking into account the flavor non-degeneracy in the sneutrino mass
spectrum.  The presence of a small relative mass splitting for the
sneutrinos, say, $ \tilde \nu _1 ,\ \tilde \nu _2 $, has the ability
to produce a third non-zero neutrino mass eigenvalue.  The
relationship between the ratio of non-zero neutrino masses and the
sneutrino mass splitting is given by~\cite{grossm04} \bea \e _ D & = &
{m_{\nu _1} \over m_{\nu _2} } \simeq 10^{-1} \D _\e ^2 , \nonumber \\
\D _\e & =& {\vert m_{\tilde \nu _1 } ^2 - m_{\tilde \nu _2 } ^2 \vert
\over 2 m_{\tilde \nu _1 } ^2 }.  \eea
\subsubsection{\bf Higgs boson decoupling} 
Our next task is to estimate semiquantitatively the Higgs sector
suppression factor $\e _H$.  Following MSSM~\cite{grossm04}, we
consider the definition \bea \e _H & = & { \vert \sum _I s^I_{ij} I_4
( m_{\tilde \nu _i} ,m_{\tilde \nu _j} ,M_{\tchi } , M_{S_I} ) - \sum
_J p^J_{ij} I_4 ( m_{\tilde \nu _i} ,m_{\tilde \nu _j} ,M_{\tchi } ,
M_{P_J}) \vert \over \sum _I \vert s^I_{ij} I_4 ( m_{\tilde \nu _i}
,m_{\tilde \nu _j} ,M_{\tchi } , M_{S_I}) \vert + \sum _J \vert p^J
_{ij} I_4 (m_{\tilde \nu _i} ,m_{\tilde \nu _j} ,M_{\tchi } , M_{P_J})
\vert }.\eea One can easily verify that $\e _H$ vanishes in the limit
of mass degenerate scalars and pseudoscalars.  The Higgs boson mass
eigenvalues $ m^2_{S_I} \ (I = 1,2,3) $ and $ m^2_{P_J}, \ (J = 1,2)
$ and mixing matrices $ U_s $ and $ U_p $, where the latter is
expressed in the basis $ (A,S)$ as $ U_p =\calr _ \g $ in terms of the
$SO(2)$ rotation matrix of angle $\g $, are determined once one 
substitutes the values of the free parameters $ \l ,\ \kappa, \ A _\l ,\
A_\kappa ,\ \tan \b = v_u/v_d $, while using the observed value $ v=
(v^2_d + v^2_u ) ^\ud = 174 $ GeV.  Ensuring a vacuum solution with
electroweak symmetry breaking at the appropriate scale, and without
tachyonic scalar modes, is known to impose strong constraints on the
NMSSM~\cite{ellis89}.  However, a systematic exploration of entire
parameter space consistent with all the physical constraints is beyond
the scope of the present work.
For a semiquantitative estimate, which is adequate for the purpose of
illustrating the typical order of magnitude values assumed by $\e _H
$, we only explore a small region of parameter space.  For this
purpose, we shall consider a modest numerical study confined to large
and small values of $x$, respectively, with the coupling constants
held fixed, where one expects to find the largest departures from the
MSSM. The mixing matrices in these two regimes are described by the
approximate formulas \bea x & >> & v_{1},\ v_{2}: \ U_s \simeq diag
(\calr _ \b , 1) \ ,\ \g \simeq \pi /2, \nonumber\\ \\ x & << &
v_{1},\ v_{2}:\ U_s \simeq \pmatrix{-\sin \a + C \cos \a & \cos \a + C
\sin \a & 0 \cr \cos \a + C \sin \a & \sin \a - C \cos \a & 0 \cr 0 &
0 & 1} , \ \g \simeq 0, \eea \\ where $ C= {2 \l A_\l x \cos ( 2 \a )
\sin 2 (\b -\a ) / [m_Z ^2 \sin (4 \b ) ] } .$

We shall study the dependence of the function $\e _H$ on various
parameters by means of two different prescriptions. In the first, we
set the dimensionless coupling constants at the renormalization group
infrared fixed point values, $\l = 0.87 , \ \kappa= 0.63 $, with the
soft supersymmetry breaking parameter $ A_\kappa$ having the fixed
value $ A _\kappa =200 \ GeV$, and vary  the parameter $ A_\l
$. This results in the variation of the mass $ m_C$ of the charged
Higgs boson, $ C^+ = \cos \b H_u^+ + \sin \b H_d^{-\star }$, which is
given by the tree level formula $ m_C^2= m_W^2 -\l ^2 (v_d^2 +v_u^2) +
2\l (A_\l +\kappa x) x /\sin (2\b ) $.  The variation of $\e _H$ with
$ m_C$ is examined at discrete values of $x$ and $\tan \b$, while the
sensitivity of these results with respect to the other fixed
parameters is examined by considering small variations around the
above reference values of the parameters.  In the second prescription,
we examine the dependence of $\e _H$ on $x$ for the choice of fixed
parameter values $ \l =0.5,\ \kappa = 0.5 ,\ A_\kappa = 100 \ GeV,\
A_\l = 100 \ $ GeV.  In both prescriptions, we assign definite mass
values to the lowest lying neutralino and the pair of lowest lying
sneutrinos, namely, $ m_{\tchi _l} = 300 $ GeV and $ m_{\tilde \nu _1}
=100 $ GeV, $ m_{\tilde \nu _2} =200 $ GeV.
while noting that the loop momentum integral $ I_4$ depends very
weakly on the input masses.

\begin{center} \begin{figure}
\epsfxsize =6. in
\epsffile{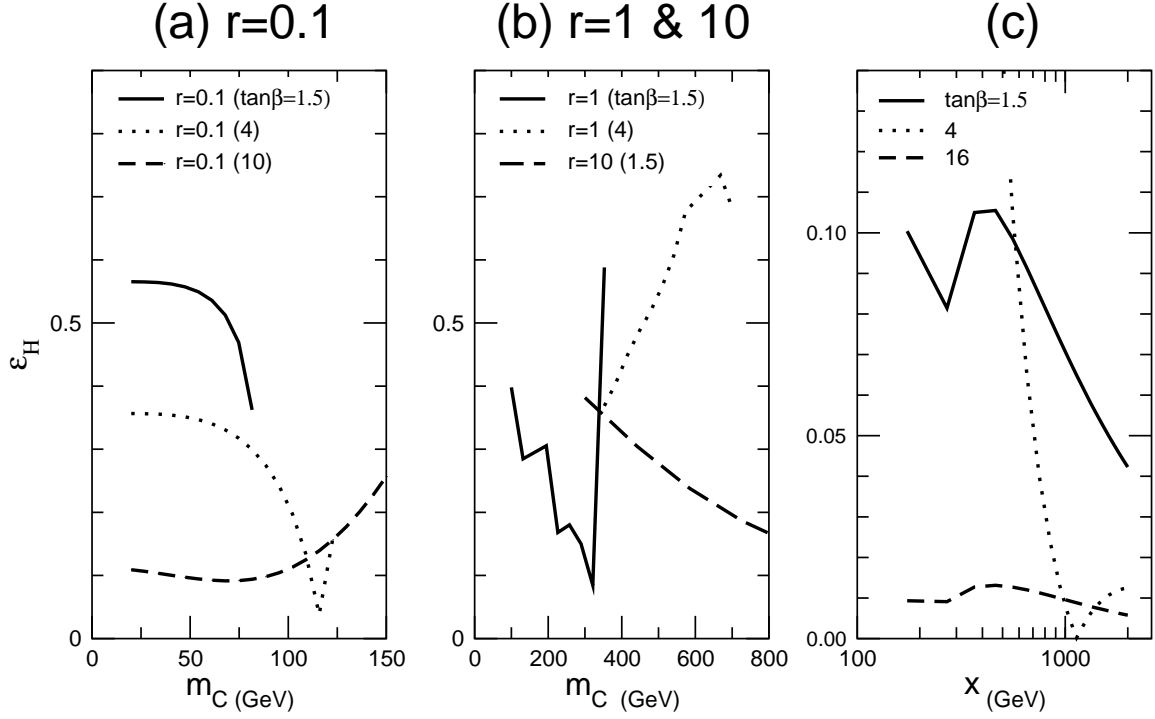}
\caption{The Higgs sector decoupling ratio $\e _H$ in the double mass
insertion approximation at one-loop order is plotted as a function of
the charged Higgs boson mass $ m_C$ for the three regimes of the VEV
ratio parameter, $ r \equiv {x\over v} = 0.1$ and $ 1 , \ 10$ in the
frames $(a) $ and $(b)$, respectively, and as a function of the $S$
field VEV, $x$, in the frame $(c) $.  The results in frames $(a) , \
(b) $ are obtained with the fixed values of parameters $ \l =0.87,
\kappa = 0.63 ,\ A_\kappa = 200 $ GeV, and with a variable parameter
$A_\l $, which determines the charged Higgs boson mass $ m_C$.
The plots in frame $(a)$ are for $\tan \b = 1.5,\ 4,\ 10 $, with $ r
=0.1 $, and those in frame $(b)$ for $ r=1,\ \tan \b = 1.5,\ 4, $ and
for $ r=10,\ \tan \b = 1.5 $, as indicated in the legends.  The curves
in frame $(c)$ are for $\tan \b = 1.5,\ 4,\ 16 $ at the fixed values
of parameters, $ \l =0.5,\ \kappa = 0.5 ,\ A_\kappa = 100 \ GeV,\ A_\l
= 100 \ $ GeV, as indicated in the legend.
The variables in the 4-point amplitude $I_4$ are chosen by setting the
masses of the lightest pair of sneutrinos at $ m_{\tilde \nu _1} =100
\ GeV ,\ m_{\tilde \nu _2} =200 \ GeV $, and that of the lightest
neutralino at $m_{\tilde \chi _1} =300 \ GeV $.}
\label{nhiggs} \end{figure} \end{center}

The plots of the ratio $\e _H$ as a function of $ m_C$ and $x$ is
displayed in Fig.~\ref{nhiggs} in the frames $ (a), (b)$ and $(c)$ for
the above two prescriptions.
For the first prescription using variable $ m_C$, the plots are
restricted to the physically acceptable values  of the parameter
$A_\l $ in which no tachyonic scalars or pseudoscalars are present in
the neutral Higgs boson sector.  In the small $x$ regime with $r
\equiv 0.1$, the lowest lying Higgs boson mass lies in the interval $
m_h \sim 50 \ - \ 20 $ GeV for $ m_C \sim 20 \ - \ 100 $ GeV, which is
excluded by the experimental limits.  In the intermediate $x$ regime
with $r=1$ and $r=10$, it is pushed up to the interval $ m_h\sim 70 \
- \ 120 $ GeV and $ m_h\sim 130 \ - \ 140 $ GeV, respectively for $
m_C \sim 100 \ - \ 300 $ GeV and $ m_C \sim 300 \ - \ 2000 $ GeV.  The
plots in the frames $(a),\ (b) $ show that the variation of $\e _H$
with $ m_C$ is slow except when one approaches the boundaries where
tachyons appear.  The typical size of the ratio is $ \e _H =
O(10^{-1})$, irrespective of the small or large values of $x$, but
decreases by a factor $2\ -\ 3$ with increasing $\tan \b $.  The
discontinuous behavior of the curves for $\e _H$ is explained by the
fact that this is the absolute value of the difference of two
amplitudes.  The plot in frame $(c)$ shows that $\e _H$ has a strong
variation with increasing $x$ in the interval $x >v $ with a typical
size $O(10^{-1})$, decreasing by a factor $ 10 $ with increasing $\tan
\b $.

We have also examined how the ratio $\e _H$ varies with small
variations about the reference values of the couplings for the first
prescription. Changing $ A _\kappa $ has a mild influence on the Higgs
boson mass spectrum and hence on $\e _H$.  Indeed, increasing $ A
_\kappa $ by a factor $ 2 - 3 $ does not affect the prediction for $\e
_H$ significantly.  Decreasing $\l $ by a factor $ 2 $ reduces $\e _H$
mildly at small $x$ and more strongly, by factors of $ 2\ -\ 5$, at
large values of the parameter $x$. A similar but weaker decrease
applies when we reduce $\kappa$ by a factor $ 2 $.  As shown by
Fig.~\ref{nhiggs}, $\e _H$ decreases rapidly with increasing $\tan \b
$, but undergoes very small changes when we allow for large variations
of $ m_\tchi , \ m_{\tilde \nu _i}$.

Thus, the main conclusion of our analysis is that the suppression
factor $\e _H$ arising from the Higgs sector is typically of order $
10^{-1} \ -\ 10^{-2}$.  This is larger than the value obtained for the
corresponding factor in the MSSM~\cite{grossm04}, $\e _H \simeq
10^{-2}\ - \ 10^{-3} $, at the values of parameters consistent with
physical constraints.  It is, however, possible that there are regions
of parameter space where $\e _H$ is significantly smaller in the
NMSSM.

\section{Flavor Symmetries} 
\label{sechoriz}

There are too many parameters in supersymmetric models, including the
nonminimal supersymmetric model, to make any specific predictions for
the neutrino spectrum. It is even difficult to identify important
contributions to the neutrino masses. Here we shall study a specific
framework, that of an Abelian flavor~(horizontal)
symmetry~\cite{froggat, nir}, where specific predictions can be made.
Flavor symmetries are usually invoked to explain the pattern of
fermion masses. However, any theory of fermion masses must also
explain why the violations of $R$-parity~(or lepton and baryon number)
are small.  This applies particularly to NMSSM with lepton number
violation coming from a trilinear type superpotential coupling that we
are considering here as the origin of neutrino masses.

We start by recalling the salient fatures of the Abelian flavor
symmetry framework. The basic idea is to use an Abelian horizontal
symmetry $U(1)_F$ to forbid most of the Yukawa couplings except
perhaps the third generation couplings. The hierarchies of fermion
masses and mixing are then generated through higher dimensional
operators involving one or more electroweak singlet scalar
fields. These fields acquire vacuum expectation values at some high
scale and give rise to the usual Yukawa couplings. More specifically,
if $\Theta$ is some such field which has charge $-1$ under $U(1)_F$,
then $X$-charge allows the non-renormalizable term in the
superpotential \be \lambda_{ij}\Phi_i \Phi_j H \left( \frac{\Theta}{M}
\right)^{n_{ij}}, \ee where $\Phi_i$ is a matter superfield of flavor
$i$,  and $H$ is a Higgs superfield with appropriate transformation
properties under the gauge group. The coupling $\lambda_{ij}$ is of
order unity, and $M$ is some large mass scale. The positive rational
numbers $n_{ij}$ are nothing but the sum of $X$-charges of $\Phi_i$,
$\Phi_j$ and $H$: \be n_{ij} = \phi_i + \phi_j + h.  \ee When $\Theta$
gets a vacuum expectation value, an effective Yukawa coupling \bea
Y_{ij} & = & \lambda_{ij} \left( \frac{<\Theta>}{M} \right)^{n_{ij}}
\nonumber \\ & \equiv & \lambda_{ij} \theta _C^{n_{ij}}, \eea is
generated. If $\theta _C $ is a small number, and if the $U(1)_F$
charges are sufficiently diverse, one can implement various hierarchies
of fermion masses and mixing. This can then be viewed as an effective
low energy theory that originates from the supersymmetric version of
the Froggatt-Nielsen mechanism at higher energies. From above we then
have the following consequences: \\ (i) Terms in the superpotential
that carry charge $n \ge 0$ are suppressed by ${\cal O}(\theta _C
^n)$, whereas those which have $n < 0$ are forbidden by the holomorphy
of the superpotential.\\ (ii) Soft supersymmetry breaking terms that
carry a charge $n$ under $U(1)_X$ are suppressed by ${\cal O}(\theta
_C ^{|n|})$.

Applying the above scheme to the neutrino mass matrix, we see that the
additive separable structure of the combined tree and loop level
contributions give us the ability to account for moderate flavor
hierarchies.  Let us first recall that the fit to the neutrino
oscillation experimental data, assuming a mass spectrum of normal kind
with mild hierarchies, favors the following approximate solution for
the three masses and mixing angles~\cite{moha04}: $ m_{\nu _3} \sim 10
^{-1} \ eV, \ m_{\nu _2} \sim 10 ^{-2} \ eV,\ m_{\nu _1} \sim 10 ^{-3}
\ eV $ and $\sin ^2 \t _{23} \sim {1\over 2} ,\ \sin ^2 \t _{12} \sim
{1\over 3} , \ \sin ^2 \t _{13} < 1.4 \ 10^{-2} $.  Of course, the
contributions that we have discussed so far are controlled by $O(100)
$ GeV weak interaction scale, which lies considerably higher than the
observed neutrino mass  scales.  Having identified the supposedly
dominant contributions, it is now necessary to find a plausible
suppression mechanism which accounts for the wide $ O(10^{12})$
hierarchy in mass scales. As in the familiar Froggatt-Nielsen
approach~\cite{froggat,nir}, we can adjust the overall size of the
contributions without an excessive fine tuning of the free parameters
by postulating that the superpotential and supersymmetry breaking
couplings of the NMSSM arise from non-renormalizable operators with
effective couplings weighted by powers of the small parameter $\t _C =
<\Theta> / M$, which we shall identify here with the Cabibbo
angle parameter, $\t _C \simeq 0.2$.  With $ h (L_i),\
h(H_{d,u}),\cdots $  denoting the Abelian
horizontal group $U(1) _F$ charges assigned to the various
superfields, one finds, $\tilde \l _i =\t _C ^{ [ h (L_i) + h(S) +
h(H_u)] } <\tilde \l _i > $, and similarly for the associated
supersymmetry breaking parameters, $ A _{\tilde \l _i } \tilde \l _i
$, with the expectation that $<\tilde \l _i^2> = O(1) ,\ < A _ {\tilde
\l _i } ^2 \tilde \l _i^2> = O(1) $.  Using the results of
subsections~\ref{sectree} and \ref{secloop}, one can write the
predicted finite neutrino mass eigenvalues as \bea && m _{\nu_3}
\simeq { x^2 \cos ^2 \b \over \tilde m _0 } < \tilde \l ^2 > \t _C ^{2
h(L_3)} ,\ {m _{\nu_3} \over m_{\nu_2} } \simeq {\cos ^4 \b \over \e
_L \e _H } {<\tilde \l ^2 > \over < A_{\l } ^2 \l ^2 > } \t _C ^{2
[h(L_3) -h(L_2)] } ,\cr && {m _{\nu_1} \over m_{\nu_2} } \simeq \e _D
\t _C ^{2 [h(L_2) -h(L_1) ] } .\eea 
In order to obtain $ m_{\nu _3} \sim 10^{-1}$,
we must have  $ \t _C ^{2 h(L_3)} \geq 10 ^{-12} $, and hence $ h(L_3) \leq
9$.  Similarly, in order to obtain $ m_{\nu _2} \sim 10^{-2} $ yields
$ h(L_3) - h(L_2) \leq 2$. Furthermore, $\ m_{\nu _1} \sim 10^{-3} $
can be achieved by lifting the mass degeneracy between sneutrinos with
$ h(L_2) \sim h(L_1)$.  Recalling the predictions for the lepton
flavor mixing angles, $\sin \t _{ij} \sim \t _C ^{h(L_i) -h(L_j) } $,
it follows that, as in the case of MSSM case~\cite{grossm04}, the
selection of horizontal symmetries involving nearly equal horizontal
charges $h(L_i) $ introduces a fine tuning problem in order to account
for the small observed mixing angle $ \t_{13}$.

\section{Summary and Conclusions}
\label {conclu}

We have studied the nonminimal supersymmetric standard model with
lepton number violation in detail. This model has a unique trilinear
lepton number violating term in its superpotential, and a corresponding
soft susy breaking scalar trilinear coupling. We have attempted   
to justify these terms on the basis of a gauged discrete  symmetry. We have
shown that these terms give a viable description of the light neutrino
Majorana mass matrix provided one stabilizes the large mass hierarchy
with respect to the weak  gauge interactions scale by invoking horizontal 
flavor symmetries.
A satisfactory feature of this extended version of the NMSSM is that the
suppressed interactions are all associated with effectively
renormalizable and dimensionless Yukawa couplings.  This mechanism
represents an economic alternative option to the familiar see-saw
mechanism of generating the light neutrino mass matrix.  Although
qualitatively similar to the bilinear lepton number violation that
occurs in MSSM, it distinctly differs from it on important
quantitative grounds. We find that only one of the three neutrinos
obtains mass at the tree level and that this has a finite component of the
massive singlet fermion.  We have also calculated the one-loop
radiative corrections to the neutrino mass matrix generated  by the
coupling of sneutrino and Higgs boson sectors. These can contribute
finite masses to the other two neutrinos in a manner favoring mild
hierarchies of normal kind for the neutrino mass spectrum  along 
with large  lepton flavor  mixing angles.  
One can reproduce a single small mixing angle,
as needed for agreement with the current experimental data,  
at the cost of a small fine tuning.

In an effort to put  on a firmer theoretical basis the different 
versions  of the NMSSM   with
renormalizable $B$ or $L$  number violation, we have
also studied the  four main gauged  $Z_N$ cyclic group 
(ordinary and R, free  and GS anomalous) realizations of the generalized
baryon, lepton and matter parities.  The constraints from the anomaly 
cancellation conditions  are so strong that no solutions exist if one 
restricts to the strictly minimal  matter field content.  However,  making the 
reasonable choice of retaining  only the least model dependent conditions,  
associated with the mixed gauge anomalies, and of admitting at least one 
extra gauge singlet chiral superfield (in addition to the standard one which 
couples to the Higgs bosons), we find interesting restricted classes of 
symmetry solutions at low   cyclic group orders $N$.

\section{Acknowledgements}
\label{acknow}
One of the authors~(PNP) would like to thank the theoretical physics group
at Saclay for its hospitality during the initial stages of this work. He
would also like to thank the Inter-University Center for Astronomy and 
Astrophysics, Pune for its hospitality while this work was being completed.
The work of PNP is supported by the University Grants Commission, India.

\appendix
\section{Classification of  cyclic discrete gauge symmetries}
\label{appex1}

The interest in generalized parities for the MSSM was historically
motivated by the need to suppress the dimension-$5$ baryon and lepton
number violating supersymmetric operators~\cite{iban92,iban95}.  Our
purpose in the present appendix is rather  to classify the discrete cyclic
group symmetries which protect the structure of renormalizable
versions of the NMSSM superpotential with baryon or lepton number
violation.  The issue of adding gauge singlet chiral supermultiplets
was considered by Lola and Ross~\cite{lola93}, although their work was
focused on applications involving non-renormalizable couplings.  Our
present treatment of this problem also slightly deviates in certain
technical details from that followed in this earlier work.

We wish to prove the existence of cyclic symmetries $Z_N$ of general
order $N$ which leave invariant the interaction superpotential of the
NMSSM with $B$ or $L$ number violation with trilinear Yukawa
interactions.  Based on the approach initiated by Ib\'a\~nez and
Ross~\cite{iban92,iban95}, one distinguishes three cases of discrete
symmetries designated as generalized baryon~(GBP), lepton~(GLP) and
matter parities~(GMP), respectively.  For each case, there are four
different realizations depending on whether the symmetry is ordinary
or R like, and whether it is Green-Schwarz (GS) anomaly free or
anomalous.  We discuss first the general classification of the
different discrete symmetries and next the consistency conditions
imposed by the cancellation of anomalies. Our considerations will be
restricted to the flavor blind symmetries.

\subsection{Ordinary symmetries}

Let us start with the ordinary anomaly free symmetries.  Recall first
 that the quark and lepton generation independent Abelian charges
 conserved by the renormalizable R parity conserving (RPC)
 superpotential couplings of the MSSM form a vector space generated by
 three continuous $U(1)$ symmetries.  A convenient basis for the  three
 independent charges is given by $\hat R , \ \hat A , \ \hat L $,
 where $\hat R =T_{3R}$ and $\hat A =Y_A $ identify with the Cartan
 generators of the right symmetry group $SU(2)_R$ and the $SU(2)$
 group embedded in $ SU(6)\times SU(2) \subset E_6 $, and $-\hat L$
 identifies with the usual lepton number.  The charges $\hat R,\ \hat
 A, \ \hat L$ assigned to quarks, leptons and Higgs boson superfields
 are displayed in the following table, along with those assigned to
 the singlet superfield $S$i, which are denoted by $ x, y, z$.  (The charge
 $\tilde P$ will appear in the next subsection in the discussion of R
 symmetries.)

\vskip 0.5 cm
\begin{center}  
\begin{tabular}{|c|cccccccc|}   \hline &&&&&&&& \\
Mode & $Q$ & $ U^c$ & $ D^c$ & $ L$ & $E^c$ & $H_d$ & $ H_u$ & $S$ \\
\hline &&&&&&&& \\ $6 Y $ & $ 1 $ &$ -4 $ &$ 2$ &$ -3 $ &$6$&$-3$&$
3$&$0$ \\ &&&&&&&& \\ $\hat R $ & $ 0$ &$ -1$ &$ 1$ &$ 0$ &$ 1$&$
-1$&$ 1$&$x$ \\ &&&&&&&& \\ $\hat A $ & $ 0$ &$ 0 $ &$ -1$ &$ -1$ &$
0$&$ 1$&$ 0$&$ y$ \\ &&&&&&&& \\ $\hat L$ & $ 0$ &$ 0 $ &$ 0$ &$ -1$
&$ 1$&$ 0$&$ 0$&$z$ \\ &&&&&&&& \\ \hline &&&&&&&& \\ $\tilde P $ & $
-1$ &$ -1 $ &$ -1$ &$ -1$ &$ -1$&$ 1$&$ 1$&$-1$ \\ &&&&&&&& \\ \hline
\end{tabular}
\end{center} 
\vskip 0.5 cm

The $Z_N ^{R} ,\ Z_N ^{A} ,\ Z_N ^{L} $ group elements are constructed
in the same way as for the continuous groups, $U(1)_{R,A,L}$, by
writing $R= e^{i\a _R \hat R},\ A= e^{i\a _A\hat A},\ L= e^{i\a _L
\hat L}$, while restricting the complex phase angles to the fixed
values, $\a _{R} = 2\pi m /N ,\ \a _{A} = 2\pi n /N ,\ \a _{L} = 2\pi
p /N $, with integer charges $ m, n, p $ defined modulo $N$.
Note that the $U(1) _{PQ} $ symmetry generated by $ g_{PQ}= R^2 A$ is
a chiral, color group anomalous symmetry which conserves all
renormalizable (RPC and RPV) trilinear couplings of the MSSM.  The
pseudoscalar Higgs boson $A $ is the pseudo-Goldstone boson of the
$U(1) _{PQ} $ symmetry present in the limit $\mu \to 0$ where the
explicit symmetry breaking bilinear coupling $\mu H_u H_d$ is absent.

The multiplicative $Z_N$ symmetries of the renormalizable
 superpotential for the quarks, leptons and Higgs bosons may be
 parameterized in terms of the generators $ g = R^m A^n L^p = g_{PQ}
 ^n R^{m-2n} L^p ,\ [m, n, p$ integers]. The symmetry solutions
 preserving $B $ and $ L$, or $ B$ alone or $L$ alone are designated
 as generalized matter, baryon and lepton parities (GMP, GBP, GLP),
 respectively.
 Thus, aside from the regular interactions with Higgs bosons, $ Q U^c
 H_u, Q D^c H_d, L E^c H_d$, the GBP are required to forbid the
 interactions $ U^c D^cD^c$ but to allow the interactions $ L H_u,\ L
 L E^c ,\ LQ D^c $, the GLP acts  in a manner forbidding the
 lepton number violating interactions, but allowing baryon number
 violating interactions $ U^c D^cD^c$, while the GMP must forbid all 
 the matter interactions.  Noting the charges for the pure matter couplings, 
 $ \hat g (LLE ^c)= g(LQD^c)=m-2n -p, \ \hat g(U^c D^cD^c) = m-2n, $
 one finds that the discrete symmetry generators preserving the MSSM
 trilinear superpotential in the three relevant cases are given
 by~\cite{iban92,iban95} \bea && GBP:\ m-2n-p =0 ;\ m-2n \ne 0 \
 \Longrightarrow \ g_{GBP} = g_{PQ} ^n (RL)^p ,\ [p\ne 0] \cr && GLP:\
 m-2n =0 ;\ m-2n -p \ne 0 , \ne 0 \ \Longrightarrow \ g_{GLP} = g_{PQ}
 ^n L^p ,\ [p\ne 0] \cr && GMP:\ m-2n -p \ne 0 , \ m-2n \ne 0, \
 \Longrightarrow \ g_{GMP} = g_{PQ} ^n R^{m-2n} L^p ,\ [m-2n \ne 0 ,\
 p \ne 0] . \nonumber
 \\
 \label{eqgps} \eea
A similar analysis applies in the NMSSM, with the generators for GBP,
GLP and GMP required to forbid the matter couplings violating baryon
number only ($ U^c D^cD^c$), lepton number only ($ L_i H_u S , \ LLE^c
, LQD^c$) and both combined, respectively.  The selection rules for
the allowed and forbidden $S$ field dependent trilinear couplings are
given by, $\hat g(H_d H_u S)\equiv n + \cals =0, \ \hat g(L H_u
S)\equiv m-n-p +\cals =0 ,\ \hat g(S^3)\equiv 3 \cals =0 ,\ [\cals =
mx +ny +pz ]$ and $\hat g( S) \equiv \cals \ne 0 , \ \hat g( S
^2)\equiv 2 \cals \ne 0$.  Except for the different conditions on the
integers $ (m, n, p) \in Z_N $, the cyclic symmetry generators have
the same functional form as in the MSSM, \bea && GBP:\ m-2n-p =0 ,\
m-n-p + \cals =0; \ m-2n \ne 0 \ \Longrightarrow \ g_{GBP} = g_{PQ} ^n
(RL)^p ,\cr && GLP:\ m-2n =0 ; m-2n -p \ne 0 , \ n-p + \cals \ne 0 \ ,
\ m-n-p + \cals \ne 0 \ \Longrightarrow \ g_{GLP} = g_{PQ} ^n L^p ,
\cr && GMP:\ m-2n -p \ne 0 , \ m-2n \ne 0, \ m-n-p + \cals \ne 0 \
\Longrightarrow \ g_{GMP} = g_{PQ} ^n R^{m-2n} L^p ,
\label{eqgps2}   \eea where the
three symmetry cases must satisfy $p\ne 0 $ along with the conditions:
$ n+\cals =0 ,\ 3 \cals =0; n\ne 0, \cals \ne 0, 2 \cals \ne 0 $.

Having classified the cyclic groups, we now wish to implement the
condition that these belong to gauged symmetries. This means that the
total contributions to the quantum anomalies from massless fermions of
the low energy theory must either vanish or be compensated by those of
the massive fermions of the high energy theory which decouple by
acquiring large Dirac or Majorana masses.  The coefficients of mixed
gauge and gravitational anomaly operators $ F _a\tilde F_a ,\ R \tilde
R ,\ F_g ^2 F_Y$ and of chiral anomaly operator, $ F_g ^3 $, acquire
the following contributions from massless fermions, \bea && \cala (
Z_N \times G_a ^2 ) =\sum _i \mu _a (\psi _i) \hat g(\psi _i) ,\ \cala
(Z_N \times grav ^2 ) =\sum _i \hat g(\psi _i) , \cr && \cala (
Z_N^3)= \sum _i \hat g ^3(\psi _i) , \ \cala ( Z_N ^2 \times U(1) _Y )
=\sum _i \hat g ^2 (\psi _i) Y_i , \eea where $\mu _a (\psi _i) $
denotes the Dynkin index of the fermion $\psi _i$ representation with
respect to gauge group factors $G_a$ of the SM gauge group $SU(3)_C
\times SU(2)_L \times U(1)_Y$ and the label $ `\text{grav}' $ stands
for the gravitational field source.  The Abelian gauge anomaly $\cala
_1 $ will be evaluated by setting conventionally the hypercharge
normalization such that, $Y(e^c)=1$, which implies that the trace of
$Y^2$ over a single quark and lepton generation amounts to $Trace(
Y^2) \equiv 2 k_1 = 10/3$.  The anomaly cancellation conditions for
the NMSSM are given by the formulae
\bea && \cala_3= \cala (SU(3)^2 \times Z_N )= -nN_g = r N , \cr && \cr
&& \cala_2= \cala (SU(2)^2 \times Z_N ) = -N_g (n+p) + N_{2h} n = r N
, \cr && \cr && \cala_1= \cala (U(1)_Y^2 \times Z_N ) = N_g(-{5n\over
6} +{p\over 2} ) + N_{2h} {n \over 2} = r N , \ \cr && \cr && \cala
_{Z^2}= \cala (U(1)_Y \times Z_N ^2 ) = -2 N_g( 2 m n + p(n-m) ) -
N_{2h}n(n-2m)
= r N , \cr && \cala _{grav}= \cala (grav ^2 \times Z_N )= -N_g
(5n-m+p) + 2 N_{2h}n +\cals
= rN +\eta s {N \over 2} , \cr && \cala _{Z^3}= \cala (Z_N ^3 )= N_g
[-3 \bigg (m^3+(n-m)^3 \bigg ) -2(n+p)^3 +(m+p)^3 ] \cr && + N_{2h}
[(n-m)^3 +m^3] + \cals ^3
= r N + \eta s {N ^3 \over 8} . \eea Here, $N_g$ denotes the number of
quark and lepton generations, $ N_{2h} $ the number of $H_d,\ H_u$
Higgs boson supermultiplet pairs, and the symbols $r ,\ s \in Z $
denote arbitrary integers (taking independent values for the different
anomalies) so that the different equations are understood to be
satisfied modulo $N$.  The additional vanishing conditions associated
with the parameter $\eta = 0,\ 1 $ for $N$ odd and even, respectively,
are introduced to account in the even $N$ case for the presence of
massive Majorana fermions in real representations of the gauge group
factor $G_a$.  The gauge singlet charges enter only through the linear
combination, $ \cals \equiv mx + ny +pz $, which is set to $ \cals =
-n$.


It is straightforward to generalize the above results to the case
involving several gauge singlet chiral supermultiplets, $S_i$.  One
just needs to assign $S_i$ the $ \hat R,\ \hat A,\ \hat L$ charges $
x_i, y_i, z_i$, and to replace in the anomaly coefficients, $ \cals
\to \sum _i \cals _i , \ \cals ^3 \to \sum _i \cals _i ^3 $.  These
additional contributions set conditions on the charges $ x_i, y_i,
z_i$ expressing the net cancellation of the anomaly coefficients $
\cala _{grav}, \ \cala _{Z^3} $.  Each allowed coupling must also be
accompanied by an additional constraint equation expressing the
associated selection rule.

The following two-stage procedure may be used in solving the anomaly
cancellation conditions for each fixed $N$ generator.  One first scans
through the non-vanishing integers $ m, \ n,\ p$ to select those
satisfying the above set of equations and next scans the non-vanishing
integers $ x, y, z \in Z _N$ which solve the equations $ \cals \equiv
mx + ny +pz = -n \ne 0, \ 2\cals \ne 0 $.  The search is most easily
implemented with the help of a numerical computer program.

We study next the Green-Schwarz anomalous discrete symmetries.  This
case differs from the anomaly free one in that the anomaly
coefficients in the effective action is now allowed to take finite
values, provided only that these are canceled by the additive
contributions to the anomalies of universal form associated with the
gauge and gravitational couplings of the model independent
axion-dilaton chiral supermultiplet.  The modified anomaly
cancellation conditions are then expressed in terms of the shifted
anomaly coefficients, $\cala _a -2k_a \d _{GS}=0,\ \cala _{grav} - 2 k
_{grav} \d _{GS}=0 , \ [a=3,2,1]$ where $ k_a $ are rational
parameters (integer quantized for non-Abelian  group factors $G_a$), $
k _{grav} =12 $,
and $ \d _{GS} $ denotes a universal model-dependent parameter
reflecting the underlying high energy theory. These conditions can
also be represented by the proportionality relations, $ \cala _3 /k_3
= \cala _2 /k_2 =\cala _1 /k_1 =\cala _{grav} /12 $. The parameters $
k_a $ in the minimal gauged unified theories are set at the numerical
values, $ k_3=k_2=1, k_1 = 5/3$.

\subsection{R symmetries}

We shall continue using the abbreviations GBP, GLP, GMP for the
generalized baryon, lepton and matter R parity discrete symmetries.  A
convenient representation of the $Z_N$ group generators can be
constructed by introducing the fermionic generator $ \tilde P $
defined by its action on the superspace differential, $\tilde P \cdot
d \t = e ^{-2i\pi /N } d \t $ and by the charge assignments of the
gauginos, $\tilde P (\tilde g) = \tilde P (\tilde W ) = \tilde P
(\tilde B) = 1 $, and of the matter and Higgs boson superfields, as
displayed in the table placed at the beginning of
Section~\ref{appex1}.  With the understanding that the charge
assignments displayed in the table for $ \tilde P $ and for $ \hat R,
\ \hat A, \ \hat L $ apply to the fermion field component of the
chiral superfields, the $Z_N$ group generators preserving the MSSM
matter-Higgs boson trilinear superpotential are simply given by, $
\tilde g = \tilde P R^m A^n L^p $
The $\tilde g $ charges of fermion and scalar field components of
chiral superfields $\psi ,\ \phi $ are then related in the usual way,
$ \hat {\tilde g} (\psi ) = \hat {\tilde g} (\phi ) -1$, so that a
superpotential term $W$ is conserved to the extent that it obeys the
selection rule, $ \tilde g ( W ) = e ^{4i\pi /N } W$, corresponding to
an R-charge of $2$.  Thus, the invariance requirement of an order $M$
superpotential monomial, $ W=\prod _ {I=1}^M \Phi _I $, can be
expressed by the condition, \bea && \hat {\tilde g} (W)= \hat {\tilde
g} (\prod _ {I=1}^M \Phi _I ) = \sum _{I=1}^M \hat {\tilde g} ( \psi
_I ) + M = 2 ,\eea implying the selection rule, $\sum _{I=1}^M \hat
{\tilde g} ( \psi _I ) =2-M$.

Applying the above discussion to the $S$ field dependent couplings,
one derives the following selection rules, valid for the three
symmetry cases: $3 \cals -2=0,\ n+2 + \cals =0; \ n+2\ne 0,\ \cals -2
\ne 0, \ 2 \cals -2 \ne 0 $.  It is again useful to single out the R
like Peccei-Quinn symmetry, $ \tilde g ^n_{PQ} = \tilde P (R^2A)^n R^2
$. This conserves all couplings with the exception of those involving
the pair of Higgs boson superfields, for which one has the selection
rules, $ \tilde g _{PQ} ^n(H_dH_u)= 2+n, \ \tilde g _{PQ} ^n(H_dH_u
S)= \tilde g _{PQ} ^n (LH_u S)= n+1 +\cals .$ Focusing, for
definiteness, on the GLP, one obtains the following selection rules
for the bilinear and trilinear couplings: $\hat {\tilde g} (H_d H_u )
= n+2 \ne 0 ,\ \hat {\tilde g} (L H_u ) = m-n-p \ne 0 , \ \hat {\tilde
g} (LLE^c)= \hat {\tilde g} (LQD^c) = m-2n-p-3 = -1 ,\ \hat {\tilde g}
(U^cD^cD^c)= m-2n-3 \ne -1 ,\ \hat {\tilde g} (H_dH_u S) = \cals + n+1
=-1 ,\ \hat {\tilde g} (LH_u S) =\cals + m-n -p -1 =-1 ,\ \tilde g (S)
= \cals -1 \ne 1,\ \hat {\tilde g} (S^2) = 2 \cals -2 \ne 0,\ \hat
{\tilde g} (S^3) = 3 \cals -3 = -1$.

 We can summarize the defining conditions for the GBP, GLP and GMP
generators and the resulting representations of the generators by the
formulas, \bea && GBP:\ m-2n-p -2 =0 ;\ m -n-p +\cals \ne 0,\ m-2n
-2\ne 0 \ \Longrightarrow \ \tilde g_{GBP} =\tilde g_{PQ} (RL)^p \cr
&& GLP:\ m-2n -2 =0 ,\ m -n-p +\cals = 0;\ m-2n -p-2\ne 0 \
\Longrightarrow \ \tilde g_{GLP} =\tilde g_{PQ} (L)^p ,\cr && GMP:\
m-2n -p-2\ne 0 ,\ m-2n -2 \ne 0 , m -n-p +\cals \ne 0 \
\Longrightarrow \ \tilde g_{GMP} =\tilde g_{PQ} R ^{m-2n-2} L^p,
\nonumber
\\
\eea
which must be complemented by the conditions $ p\ne 0 $ and $n+2
+\cals =0$.  The anomaly cancellation conditions are now readily
evaluated by inspection of the table given at the beginning of
Section~\ref{appex1} which displays the fermion modes charges.  One
must include in the mixed gauge anomalies the contributions from the
spin $1/2 $ gauginos, and in the gravitational anomaly those from the
gauginos and the spin $3/2 $ gravitinos which add to the anomaly
coefficient $1$ and $ -21 $ per mode, respectively.  The contribution
from the SM gauge group gauginos amounts then to $\sum _a dim (G_a ) =
12$.  The anomaly coefficients for the gauge, gravitational and chiral
anomalies are given by the formulas \bea && \cala_3= \cala (SU(3)^2
\times Z_N )= 6 -N_g(4+n) = r N, \cr && \cr && \cala_2= \cala (SU(2)^2
\times Z_N ) = 4 -N_g (4+n+p) + N_{2h} (n + 2) = r N , \cr && \cr &&
\cala_1= \cala (U(1)_Y^2 \times Z_N ) = N_g (-{10\over 3} -{5n\over 6}
+{p\over 2} ) + N_{2h} ({n\over 2} +1 ) = r N , \cr && \cr && \cala
_{Z^2}= \cala (U(1)_Y \times Z_N ^2 ) \cr && = N_g [ 1 -2(1+m)^2 +
(1-m+n)^2 - (1+n+p)^2 + (-1+m+p)^2 ] \cr && + N_{2h} [ (1-m+n)^2
-(1+m)^2] = r N , \cr && \cr && \cala _{grav}= \cala (grav ^2 \times
Z_N )= N_g (-15 -5n +m-p ) + N_{2h} (4 +2n) \cr && + 12 -21 + (-1
+\cals ) = rN +\eta s {N \over 2} , \cr && \cr && \cala _{Z^3}= \cala
(Z_N ^3 )
= N_g [ -6 -3(1+m)^3 -3 (1+n-m)^3 -2 (1+n+p)^3 + (-1+m+p)^3 ] \cr &&
+2 N_{2h} [ (1+n-m)^3 +(1+m)^3 ] + (-1 +\cals )^3 = r N + \eta s {N ^3
\over 8} . \eea The selection rule, $(-1+\cals ) \equiv (-1+mx+ny+pz)
= -(n+3)$, may be used to remove the explicit dependence of the
anomaly coefficients on the singlet field charges.  The case of
anomalous GS symmetries is analyzed in the same way as for the
ordinary symmetries by introducing the shifted anomaly coefficients.
The search of generalized parity solutions can also follow a similar
two-stage procedure as described earlier. One solves in a first stage
the anomaly cancellation equations at fixed $N$ for the integers
$(m,n,p)$, and in a second stage the selection rules $ n+2+\cals =0, \
3 \cals -2 =0, \ \cals -2 \ne 0, \ 2 \cals -2 \ne 0$ for the $S$ field
charges $(x,y,z)$.  To conclude, we note that the approach discussed
here appears more systematic than the alternative one where one solves
the anomaly cancellation equations after assigning charges to the
various particles $ \a _Q, \ \a _{U^c} , \cdots $, subject to the
selection rules.



\begin{thebibliography}{99}
\def\NPB#1#2#3 {{\rm Nucl.~Phys.}  {\bf{B#1}}, {#3} (#2)}
\def\NPA#1#2#3 {{\rm Nucl.~Phys.}  {\bf{A#1}}, {#3} (#2)}
\def\PLB#1#2#3 {{\rm Phys.~Lett.}  { {\bf{B#1}}, {#3} (#2)}}
\def\PR#1#2#3 {{\rm Phys.~Rep.}  {\bf#1}, {#3} (#2)} 
\def\PRD#1#2#3 {{\rm Phys.~Rev.}  { \bf{D#1}}, {#3} (#2)}
\def\PRL#1#2#3 {{\rm Phys.~Rev.~Lett.}  {\bf{#1}}, {#3} (#2)} 
\def\ZPC#1#2#3 {{\rm Z.~Phys.}  {\bf C#1}, {#3} (#2)} 
\def\JHEP#1#2#3 {{\rm JHEP} {\bf C#1}, {#3} (#2)} 
\def\IJMP#1#2#3 {{\rm Int. J. Mod. Phys.}  {\bf A#1}, {#3} (#2)} 
\def\JMP#1#2#3 {{\rm J. Math. Phys.}  {\bf #1}, {#3} (#2)}

\bibitem{wess} J. Wess and J. Bagger, ``Supersymmetry and
Supergravity'' (Princeton University Press, Princeton, NJ, 1992);
P. Nath, R. Arnowitt and A. H. Chamseddine, `` Applied N = 1
Supergravity'' (World Scientific, Singapore, 1984).
                                                                                
\bibitem{fayet} P. Fayet, Nucl. Phys.  {\bf B90}, 104 (1975);
R. K. Kaul and P. Majumdar, Nucl. Phys. {\bf B199}, 36 (1982);
H. P. Nilles, M. Srednicki and D. Wyler, Phys. Lett. {\bf B120} 346
(1983); J. P. Derendinger and C. Savoy, Nucl. Phys. {\bf B237}, 307
(1984).

\bibitem{ellis89} J. Ellis, J.F. Gunion, H.E.  Haber, L. Roszkowski,
and F. Zwirner, Phys. Rev. {\bf D} 39, 844 (1989).

\bibitem{drees89} M. Drees, Int. J. Mod. Phys. {\bf A4}, 3635 (1989).

\bibitem{nmssm1} P. N. Pandita, Phys. Lett. {\bf B318}, 338 (1993);
Z. Phys. {\bf C59}, 575 (1993); U. Ellwanger, Phys. Lett. {\bf B303},
271 (1993); U. Ellwanger, M. Rausch de Traubenberg and C. A. Savoy,
Phys. Lett. {\bf B315}, 331 (1993); S.F. King and P. L. White,
Phys. Rev. {\bf D} 53, 4049 (1996).

\bibitem{nmssm2} B. Ananthanarayan and P.N. Pandita,
Int. J. Mod. Phys. {\bf A12}, 2321 (1997); Phys. Lett. {\bf B371}, 245
(1996); Phys. Lett. {\bf B353}, 70 (1995).
 

                                                                                
\bibitem{nmssmft} R. Dermisek and J. F. Gunion, Phys. Rev. Lett. {\bf
95} 041801 (2005), hep-ph/0502105; hep-ph/0510322.
                                                           
\bibitem{dob00} B.A. Dobrescu and K.T. Matchev, \JHEP{0009}{2000}{031}
, hep-ph/0008192;
B.A. Dobrescu, G. Landsberg, and K.T. Matchev, \PRD{63}{2001}{075003}
, hep-ph/0005308.


\bibitem{ellwang04} U. Ellwanger, J.F. Gunion, and C. Hugonie,
\JHEP{0507}{2005}{04} , hep-ph/0503203.

\bibitem{choi04} P. N. Pandita, Phys. Rev. D50, 571 (1994);
P. N. Pandita, Z. Phys. C63, 659 (1994); S.Y. Choi, D.J. Miller, and
P.M. Zerwas, \NPB{711}{2005}{83} , hep-ph/0407209.

\bibitem{moort05} G. Moortgat-Pick, S. Hesselbach, F. Franke, and
H. Fraas, hep-ph/0508313.

\bibitem{gunion05} J.F. Gunion, D. Hooper, and B. McElrath,
hep-ph/0509024.

\bibitem{weinberg} S. Weinberg, Phys. Rev. {\bf D26}, 287 (1982);
N. Sakai and T. Yanagida, Nucl. Phys. {\bf B197}, 133 (1982).
                                                                                
\bibitem{farrar} G. Farrar and P. Fayet, Phys. Lett. {\bf B76}, 575
  (1978).
                 
\bibitem{nmssmrp1} P. N. Pandita and P. Francis Paulraj,
Phys. Lett. {\bf B462}, 294 (1999).
                                                               
\bibitem{nmssmrp2} P.N. Pandita, \PRD{64}{2001}{056002} .

\bibitem{grosshaber99} Y. Grossman and H.E. Haber,
\PRD{59}{1999}{09308} .
                          
\bibitem{gross99} Y. Grossman and H.E. Haber, \PRL{78}{1997}{3438} ;
\PRL{59}{1999}{093008} .

\bibitem{haberbasis} Y. Grossman and H.E. Haber, hep-ph/0005276.

\bibitem{davidson00} S. Davidson and M. Losada, \JHEP{0005}{2000}{021}
  , hep-ph/0005080; \PRD{65}{2002}{075025} , hep-ph/0010325;
S. Davidson, M. Losada, and N. Rius, \NPB{587}{2000}{118} ,
hep-ph/9911317.

\bibitem{abada02} A. Abada, S.  Davidson and M. Losada,
\PRD{65}{2002}{075010} , hep-ph/0111332; A. Abada, G. Bhattacharyya,
and M. Losada, \PRD{66}{2002}{071701(R)} , hep-ph/0208009.
  
\bibitem{borzum02} F. Borzumati and J.S. Lee, \PRD{66}{2002}{115012} ,
hep-ph/0207184.

\bibitem{chun02} E.J. Chun, D.W. Jung, and J.D. Park,
\PLB{557}{2003}{233} , hep-ph/0211310; E.J. Chun and S.K. Kang,
\PRD{61}{2000}{075012} .

\bibitem{grossm04} Y. Grossman and S. Rakshit, \PRD{69}{2004}{093002}
, hep-ph/0311310.

\bibitem{valle03} M. A. Diaz et al., \PRD{68}{2003}{013009}
, Erratum-ibid. {\bf D71}, 059904 (2005), hep-ph/0302021, 
and references therein.

\bibitem{moha04} For a recent reappraisal of the theoretical situation
on neutrino masses and mixing see, e,g., R.N.  Mohapatra et al.,
hep-ph/0412099.

\bibitem{benak97} K. Benakli and A.Y. Smirnov, \PRL{79}{1997}{4314} .

\bibitem{chun99} E.J. Chun, \PLB{454}{1999}{304} ; E.J. Chun and
H.B. Kim, hep-ph/9906392.

\bibitem{chemtob} For an up-to-date review and references, see, e.g.,
M. Chemtob, Prog. Part. Nucl. Phys.  {\bf 54}, 71 (2005); R. Barbier
et al., hep-ph/0406039.

\bibitem{wilczek} L. Krauss and F. Wilczek, \PRL{62}{1989}{1221} ;
T. Banks and M. Dine, \PRD{45}{1991}{1424} .

\bibitem{presk91} J. Preskill, S.P. Trivedi, F. Wilczek, and
M.B. Wise, \NPB{363}{1991}{207} .
              
\bibitem{banksdine91} T. Banks and M. Dine, \PRD{45}{1991}{1424} .

                                                
\bibitem{castano} D.J. Casta\~no, D.Z. Freedman, and C. Manuel,
\NPB{461}{1996}{50} .
                                                                       
\bibitem{dreiner} H. Dreiner, Susy '95, eds. I. Antoniadis and
H. Videau, (\'Editions Fronti\`eres, Gif-sur-Yvette, 1993); A. H.
Chamseddine and H. Dreiner, \NPB{458}{1996}{65} ; \NPB{447}{1995}{195}
.

\bibitem{iban92} L.E. Ib\'a\~nez and G.G. Ross, Phys. Lett. {\bf 260},
  291 (1991); Nucl. Phys. {\bf B368}, 3 (1992).

\bibitem{iban95} L.E. Ib\'a\~nez, \NPB{218}{1982}{514} .

\bibitem{lola93} S. Lola and G.G. Ross, \PLB{314}{1993}{336} .


\bibitem{schechter82} J. Schechter and J.W.F. Valle, Phys. Rev. {\bf
D 25}, 774 (1982).

\bibitem{schechter80} J. Schechter and J.W.F. Valle, Phys. Rev. {\bf
D 21}, 309 (1980); Phys. Rev. {\bf D22}, 2227 (1980).
                                                                                

\bibitem{passvel79} G. 't Hooft and M. Veltman, Nucl. Phys.  {\bf
B153}, 365(1979); G. Passarino and M. Veltman, \NPB{160}{1979}{151} .
									       
\bibitem{froggat} C. D. Froggatt and H. B. Nielsen, Nucl. Phys. {\bf
B147}, 277 (1979).

\bibitem{nir} M. Leurer, Y. Nir and N. Seiberg, Nucl. Phys. {\bf
B398}, 319 (1993); Y. Nir and N. Seiberg, Phys. Lett. B {\bf 309}, 337
(1993); M. Leurer, Y. Nir and N. Seiberg, Nucl. Phys. {\bf B420}, 468
(1994).
 



\end{thebibliography}
\end{document}